# The VVDS-SWIRE-GALEX-CFHTLS surveys: Physical properties of galaxies at z below 1.2 from photometric data ⋆

C. J. Walcher[1,2], F. Lamareille[5,12], D. Vergani[5,6], S. Arnouts[2,3], V. Buat[2], S. Charlot[1], L. Tresse[2], O. Le Fèvre[2], M. Bolzonella[5], J. Brinchmann[4,23], L. Pozzetti[5], G. Zamorani[5], D. Bottini[6], B. Garilli[6], V. Le Brun[2], D. Maccagni[6], B. Milliard[2], R. Scaramella[7,8], M. Scodeggio[6], G. Vettolani[7], A. Zanichelli[7], C. Adami[2], S. Bardelli[5], A. Cappi[5], P. Ciliegi[5], T. Contini[12], P. Franzetti[6], S. Foucaud[13], I. Gavignaud[14], L. Guzzo[15], O. Ilbert[16], A. Iovino[15], H.J. McCracken[1,11], B. Marano[17], C. Marinoni[18], A. Mazure[2], B. Meneux[24,25], R. Merighi[5], S. Paltani[19,20], R. Pellò[12], A. Pollo[21], M. Radovich[22], E. Zucca[5], C. Lonsdale[9], and C. Martin[10]

*(Affiliations can be found after the references)*

/

**ABSTRACT**

Measuring the build-up of stellar mass is one of the main objectives of studies of galaxy evolution. Traditionally, the mass in stars and the star formation rates have been measured by different indicators, such as photometric colours, emission lines, and the UV and IR emission. We intend to show that it is possible to derive the physical parameters of galaxies from their broad-band spectral energy distribution out to a redshift of 1.2. This method has the potential to yield the physical parameters of all galaxies in a single field in a homogeneous way, thus overcoming problems with the sample size that particularly plague methods relying on spectroscopy. We use an extensive dataset, assembled in the context of the VVDS survey, which reaches from the UV to the IR and covers a sample of 84073 galaxies over an area of 0.89 deg$^2$. We also use a library of 100000 model galaxies with a wide variety of star formation histories (in particular including late bursts of star formation). We find that we can determine the physical parameters stellar mass, age, and star formation rate with good confidence. We validate the star formation rate determination in particular by comparing it to a sample of spectroscopically observed galaxies with an emission-line measurement. While the attenuation in the galaxies shows more scatter, the mean over the sample is unbiased. Metallicity, however, cannot be measured from rest-frame optical photometry alone. As a first application we use our sample to build the number density function of galaxies as a function of stellar mass, specific star formation rate, and redshift. We are then able to study whether the stellar mass function at a later time can be predicted from the stellar mass function and star formation rate distribution at an earlier time. We find that, between redshifts of 1.02 and 0.47, the predicted growth in stellar mass from star formation agrees with the observed one. However, the predicted stellar mass density for massive galaxies is lower than observed, while the mass density of intermediate mass galaxies is overpredicted. This apparent discrepancy can be explained by major and minor mergers. Indeed, when comparing with a direct measurement of the major merger rate from the VVDS survey, we find that major mergers can account for about half of the mass build-up at the massive end. Minor mergers are very likely to contribute the missing fraction.

## 1. Introduction

Traditionally, astronomers have tackled the wide variety in the appearance of galaxies by classification. The one by Hubble (1936) with extensions is still, of course, the one most used for its simplicity. At higher redshifts, classifications have been based on red vs. blue colours (e.g. Lilly et al. 1995, Bell et al. 2004), spectral type (e.g. Tresse et al. 1996), morphological information (e.g. Brinchmann et al. 1998), or principal component analysis of the spectra (e.g. De Lapparent 2003). All these classifications exploit those quantities that are directly observable in the respective surveys. Their relation to the physical properties of galaxies has made them immensely useful for galaxy evolution studies. The colour bimodality and the number density evolution of the different colour classes in particular, have provided a very useful way to study galaxy evolution from large samples. The usefulness of other classifications, such as the classification into four SED-types described and used in Zucca et al. (2006) and Tresse et al. (2007) remains underexplored.

However, it needs to be kept in mind that a classification in terms of observed quantities does not fully represent the fundamental physical differences between galaxies. The colour bimodality, for example, draws its usefulness from the notion that red galaxies are passive; i.e. the observed colour is related to the underlying physical parameter "recent star formation history" (SFH). Much work has been done to understand and quantify the contamination from star-forming, but highly extincted, galaxies to the red sequence (e.g. Wolf et al. 2005, Franzetti et al. 2007).

While classification thus has provided a framework and an important tool for galaxy studies, it is clear that galaxies cannot be divided uniquely into well-defined "species" as in Linnaean taxonomy (Linné, 1735). If anything, it is the underlying continuum of physical properties that allows to measure and understand their evolution. It is also only in terms of this continuum of physical properties that the results from different datasets become comparable.

One of the fundamental physical concepts is the SFH of galaxies. This is hard to come by in its entirety from archaeological study alone (see Panter et al. 2007, for a recent attempt on local galaxies from the Sloan Digital Sky Survey, SDSS). As a first attempt, the SFH can be parameterized in SFR[1] (star formation rate, recent SFH) and $M^*$ (stellar mass, integrated SFH). The SSFR (specific star formation rate, SSFR = SFR / $M^*$) is also often used as an alternative to the SFR. It represents the ratio of recent and past SFH of a galaxy (see also Kennicutt et al. 2003 for the b parameter). Indeed, recent studies have shown that, in the parameter space of log($M^*$) and log(SSFR), the blue cloud of the colour bimodality becomes a star-forming sequence (e.g. Brinchmann et al. 2004, Salim et al. 2004, Zheng et al. 2007, Feulner et al. 2005, Elbaz et al. 2007, Noeske et al. 2007, Wyder et al. 2007). Star formation rates for galaxies with low SFRs have been hard to quantify, as star formation rate indicators, such as emission lines and infrared (IR) flux, become hard to measure at low SFRs.

This study aims at establishing a method of deriving the physical parameters of galaxies by fitting their rest frame optical spectral energy distributions out to redshift 1.2. This method has two advantages over the complementary approach of measuring each physical parameter by its own tracer: 1) it allows physical properties to be derived in a homogeneous way for *all* galaxies in a given sample and 2) the derived parameters are all internally consistent for one specific galaxy. Finally, as the method uses only photometric data points, it has the potential to help in the exploitation of the information from future large photometric surveys.

One of the corner-stones of the present method is the trust that can be put in the stellar population model that predicts the shape of the SED and relates it to the underlying physical properties. Earlier, systematic work has concentrated mostly on galaxies in the local universe (e.g. Brinchmann & Ellis, 2000). Salim et al. (2005, 2007) exploited the comprehensive data from the SDSS and the Galex mission to demonstrate that fits of stellar population model SEDs to observed broad-band SEDs do yield results that agree with those from more classical tracers of physical properties, in particular emission lines. Johnson et al. (2007) have used Galex, SDSS and Spitzer data to further test the consistency of the predictions from stellar population models with the observed data, concentrating in particular on the effects of dust. Schawinski et al. (2007) have developed a method for recovering star formation history parameters using broadband photometry from the UV to the near-IR and combine it with information from spectral absorption indices. They show that the combination can help to break degeneracies between the parameters age, mass, dust and metallicity. Burgarella et al. (2005) have combined the Galex and IRAS surveys to build multi-wavelengths, broadband SEDs that are again compared to the predictions of a stellar population model. They particularly show that for a sample selected in the far infrared (FIR), where the galaxies are by definition dominated by dust, the attenuation estimate derived from a fit to the optical data alone can be off by as much as a factor of 100. Finally, Iglesias-Paramo et al. (2007) have extended these studies to a redshift range between 0.2 and 0.7, showing that the results obtained from SED fitting can be used to study the cosmic evolution of star formation rate and dust content. For recent studies exploiting SED fitting at high redshift to obtain stellar masses, we refer to e.g. Maraston et al. (2006), Berta et al. (2008) and references therein.

Throughout the paper we quote AB magnitudes (Oke 1974) and use a cosmology where $H_0 = 70$ km s$^{-1}$ Mpc$^{-1}$, $\Omega_{matter} = 0.3$, and $\Omega_\Lambda = 0.7$.

## 2. Determining physical parameters

We wish to determine the physical parameters, such as $M^*$ and SFR (see Table 1), for the galaxies in our sample by fitting observed SEDs with model SEDs from stellar population synthesis models.

### 2.1. Method

We first remind the reader that the spectral energy distribution (SED) of a stellar population at time $t$ can be written as

$$L_\lambda(t) = \int_0^t dt' \Psi(t-t') S_\lambda[t', \zeta(t-t')] T_\lambda(t-t'). \quad (1)$$

Here, $\Psi(t-t')$ is the instantaneous star formation rate at time $(t-t')$, $\zeta(t-t')$ is the metallicity, $T_\lambda(t-t')$ gives the fraction of $S_\lambda[t', \zeta(t-t')]$ transmitted through the interstellar medium at time $(t-t')$, and $S_\lambda[t', \zeta(t-t')]$ is the power radiated per unit wavelength per unit initial mass by a single-age stellar population of age $t'$ and metallicity $\zeta(t-t')$. For the purpose of using it later, we also introduce here the quantities $R(t')$, which is the fraction of mass in stars returned to the interstellar medium due to mass loss and supernovae explosions for a single stellar population of age $t'$, and the stellar mass to light ratio at wavelength $\lambda$, i.e. $\Upsilon_\lambda(t') = (1 - R(t'))/S_\lambda(t')$.

In principle Equation 1 can be solved directly, see e.g. Panter et al. (2007), Ocvirk et al. (2006), Walcher et al. (2006) (references can also be found in Cid-Fernandez, 2007). While this approach avoids prior assumptions on the functional form of $\Psi(t-t')$, degeneracies and problems with the accuracy of the stellar population model are difficult to control. Another classical approach, that of minimizing $\chi^2$ over a number of precomputed model SEDs, yields a single best-fit model star formation history. While in principle $\chi^2$ statistics also provide confidence regions, these can be inappropriate for two reasons in the present case: 1) unsuitable model and 2) overlooked degeneracies.

We therefore opt here for the method recently adopted by e.g. Kauffmann et al. (2003), Brinchmann et al. (2004), and Salim et al. (2007). This is based on a Bayesian understanding of probabilities and was first proposed in Kauffmann et al. (2003, see Appendix A of that paper for a rigorous treatment of the method). Put in words, we assume that

$$P(M|D) \propto P(M) \times P(D|M), \quad (2)$$

where $P(D|M)$ is the probability of the data given the model (which is what we can measure), $P(M|D)$ is the probability of the model given the data (which is what we would like to know), and $P(M)$ encodes our prior knowledge in the form of a model probability. Assuming Gaussian uncertainties, the probability, or likelihood, of the data given the model is $P(D|M) = e^{-\chi^2/2}$.

The probabilities for each model are then marginalized over all parameters except the one we want to derive, which yields a probability distribution function (PDF). The median of this distribution and its confidence regions are then used to derive a robust estimate of value and error for the parameter that is to be determined. Degeneracies are immediately visible as distorted (non-Gaussian) PDFs or large confidence regions for a particular parameter. This method is "Bayesian" because we use a *precomputed* library of model galaxies, thus introducing our prior knowledge (or belief) about the functional form of $\Psi(t - t')$ into the finally resulting probability (see Figure 1 for our prior assumptions). It is particularly important to use a suitable prior on $\Psi(t - t')$ and $\zeta(t - t')$ to avoid systematic biases. In particular, an artificially narrow distribution of star formation histories (e.g. excluding secondary bursts) will affect the results. Obviously this is the case for the resulting physical property, i.e. if the "true" parameter combination does not exist in the library. However, the *uncertainties* are also affected, as these will be underestimated if a degenerate parameter combination lacks from the library. Similar approaches to fit broad-band SEDs have been used with success in the literature, see e.g. Finlator, Dave & Oppenheimer (2007) and Blanton & Roweis (2007). Examples at higher redshifts include Sawicki & Yee (1998) for the Hubble deep field and Yan et al. (2006) who go out to redshift 6.

---

[1] For simplicity, we use throughout the paper the same abbreviation, i.e. SFR, for different determinations of the parameter "SFR" (e.g. from photometry, spectroscopy, etc.). These determinations are not necessarily equivalent, as the different tracers probe the SFR over different timescales.

The first step is thus to precompute a large library of 100000 stochastic representations of star formation histories (SFHs), with varying parameters describing the underlying SFH, dust attenuation, metallicity, and the strength and number of bursts. Secondary bursts can be produced by a number of mechanisms: accretion and cooling of gas (see e.g. Birnboim, Deckel & Neistein, 2007), close encounters (e.g. Barton et al. 2000) and mergers (e.g. Mihos & Hernquist 1996), infall into clusters (e.g. Marcillac et al. 2007) and more. It is therefore important to allow for such secondary bursts in the library of pre-computed SFHs. Note that our SFHs are parameterized functions and are not based on explicit, self-consistent modelling of galaxy evolution. Each one of these SFHs corresponds to a fictitious "model galaxy", which is the term we will use throughout this paper. Photometric properties of each model galaxy are readily derived from (red-shifted) model spectra by convolving with instrumental filter response functions.

Our ultimate goal is to derive the galaxies' physical properties to understand their evolution. When constructing our library of star formation histories, we therefore also compute such parameters, which are *not* input parameters to the computation of the library. Figure 1 shows the prior distributions of the input parameters formation time, time of the last burst, metallicity, exponent factors for the underlying SFH ($\Psi(t) \propto e^{-\gamma t}$, where $t$ is in Gyr) and attenuation $\tau_V$[2] ($T_\lambda = e^{-\tau_\lambda}$, and $\lambda$ is the effective wavelength of the V-band), as well as the prior distribution of the derived parameters mean mass-weighted age, mean light-weighted age ($r$-band), star formation rate in the mean over the last 100 Myr and stellar mass. See also Equation 1 and Table 1 for definitions of the parameters. It is important to realise that the prior on the library changes with redshift, as model galaxies that have ages older than the universe at any given redshift need to be excluded from the fitting. A library therefore needs to be carefully constructed, so the prior changes relatively little with redshift. In Figure 1 we show the prior at the two redshifts that bracket our study.

We use a preliminary version of the stellar population model described in detail in Charlot & Bruzual (2007, CB07). Although the reader is referred to this paper for details, we state some important points. 1) The model covers the full wavelength range from 900Å to 10 $\mu$m in stellar light emission for a wide range of age between 0.1 Myr and 20 Gyr and for metallicities between a fifth and twice solar. This wide coverage of parameter space is highly desirable for coping with the large dataset over a large redshift interval that we intend to study. 2) As shown by Maraston (2005, M05) and Maraston et al. (2006), it is important to correctly treat the emission from the stars that are in the thermally-pulsing asymptotic giant branch (TP-AGB) phase of stellar evolution. TP-AGB stars are especially important at a redshift of 1, where many galaxy spectra are dominated by emission from these intermediate age stars. The CB07 models include an updated treatment of this evolutionary phase from the models of Marigo & Girardi (2007). This also is the main improvement in our context over the Bruzual & Charlot (2003, BC03) model. 3) The initial mass function (IMF) is an adjustable parameter of the model and we follow BC03 in choosing a Chabrier (2003b) IMF. The spectral properties obtained using the above IMF are similar to those obtained using the Kroupa (2001) IMF. We adopt the Chabrier IMF because it is physically motivated and provides a better fit to counts of low-mass stars and brown dwarfs in the Galactic disc (Chabrier 2001, 2002, 2003a).

### 2.2. Internal consistency checks

Before applying the method to real data it is necessary to check for its internal consistency, as well as intrinsic degeneracies, in particular because we are applying it to objects with varying redshifts. To this end we carried out a set of simulations in which we attempted to recover the known properties of our galaxies from artificially downgraded SEDs. We first produce a representation of each model SED with added noise, by assuming a signal-to-noise (S/N) ratio and Gaussian errors, which we call a "pseudo-galaxy". This procedure thus concentrates on the effect of measurement errors. We then fit every pseudo-galaxy with the original model catalogue, excluding the model galaxy that was used to generate this particular pseudo-galaxy. The pseudo-galaxy SEDs are produced over the same photometric bands that will be used later on real data, i.e. NUV, B, V, R, I, u$_*$, g', r', i', z', J, K, 3.6$\mu$m, 4.5$\mu$m (see below, Section 3).

We do not attempt here to show the effects of varying observing conditions or varying data coverage. We concentrate only on the limitations that are intrinsic to the method, *even for a perfect model and 14 photometric bands*. The simulation results obtained for a constant S/N of 10 in all bands at the two ends of our redshift range are shown in Figure 2. The mean and scatter of the distributions are given in Table 2. In the case of the parameter age of the beginning of star formation T$^{form}$, the distribution of original versus fitted properties shows a general point: this parameter is not a well-defined quantity in terms of a measurement, it is an input model parameter[3]. This is why the mean light-weighted age $\langle age_r \rangle$ is usually introduced to measure the "age" of a galaxy. For $\langle age_r \rangle$ and stellar mass M$^*$, our method is known to be robust, as is confirmed by the results shown in Fig. 2. This has been exploited in several papers, some examples have been listed in the introduction. While the SSFR is also generally well-constrained, there is a minimal SSFR threshold at SSFR $\approx 10^{-12}$, under which the fitted SSFR shows a plateau. This is due to the fact that the SSFR is mainly determined by the UV-optical colours. However, these colours do not change very much with SSFR anymore when the SSFR sinks below this threshold value. This limit is due to a physical property of the SEDs and depends only very weakly on the assumed S/N ratio. Note that all age-related properties, i.e. $M^*$, $\langle age_r \rangle$, SFR and $T^{form}$ show a smaller scatter at higher redshift. This is due to the fact that the age of the universe is smaller, thus limiting the possible diversity in SFHs. This in turn reduces degeneracies. While metallicity and attenuation parameter show large scatter, it is noteworthy that they are not biased. Thus, for a large sample, mean values can be considered representative.

---

[2] For a full explanation of our dust model, see Charlot & Fall (2000).

[3] In the framework of the hierarchical model of galaxy formation it is actually also questionable, whether it has any physical meaning.

**Table 1.** Parameter definitions

| Designation (1) | Definition (2) | Range in library (3) |
|---|---|---|
| Formation time $T_{form}$ | Time since the galaxy first started forming stars | $10^8 - 14 \times 10^9$ yrs |
| Timescale of SFH $\gamma$ | $\Psi(t) \propto e^{-\gamma t}$ | $0 - 7$ Gyr$^{-1}$ |
| Stellar mass $M^{*a}$ | $M^* = \int_0^{T_{form}} dt \Psi(t)(1-R(t))$ | 0.01-10 $M_\odot$ |
| Star formation rate SFR$^a$ | $SFR = \int_0^{10^8} dt \Psi(t)/10^8 \text{yrs}$ | $10^{-30} - 10^{-7.5}$ $M_\odot$/yr |
| Time of the last burst $T_{lb}$ | Time since begin of last burst | $0 - 14 \times 10^9$ yrs |
| Metallicity Z | Abundance of all metals heavier than He | 0.2 - 2 $Z_\odot$ |
| Mean age in r-band $\langle age_r \rangle$ | $\langle age_r \rangle = \dfrac{\int_0^{T_{form}} dt \Psi(t)(1-R(t))\Upsilon_r(t)^{-1} t}{\int_0^{T_{form}} dt \Psi(t)(1-R(t))\Upsilon_r(t)^{-1}}$ | $10^6 - 14 \times 10^9$ yrs |
| Mean mass weighted age $\langle age_m \rangle$ | $\langle age_m \rangle = \dfrac{\int_0^{T_{form}} dt \Psi(t)(1-R(t))t}{\int_0^{T_{form}} dt \Psi(t)(1-R(t))}$ | $10^6 - 14 \times 10^9$ yrs |
| Effective attenuation $\tau_V$ | $T_V = e^{-\tau_V}$ | 0 - 6 |

$^a$ The parameters $M^*$ and SFR are subject to renormalization for each galaxy, depending on the ratio between the intrinsic luminosity of the object and the intrinsic luminosity of the model galaxy.

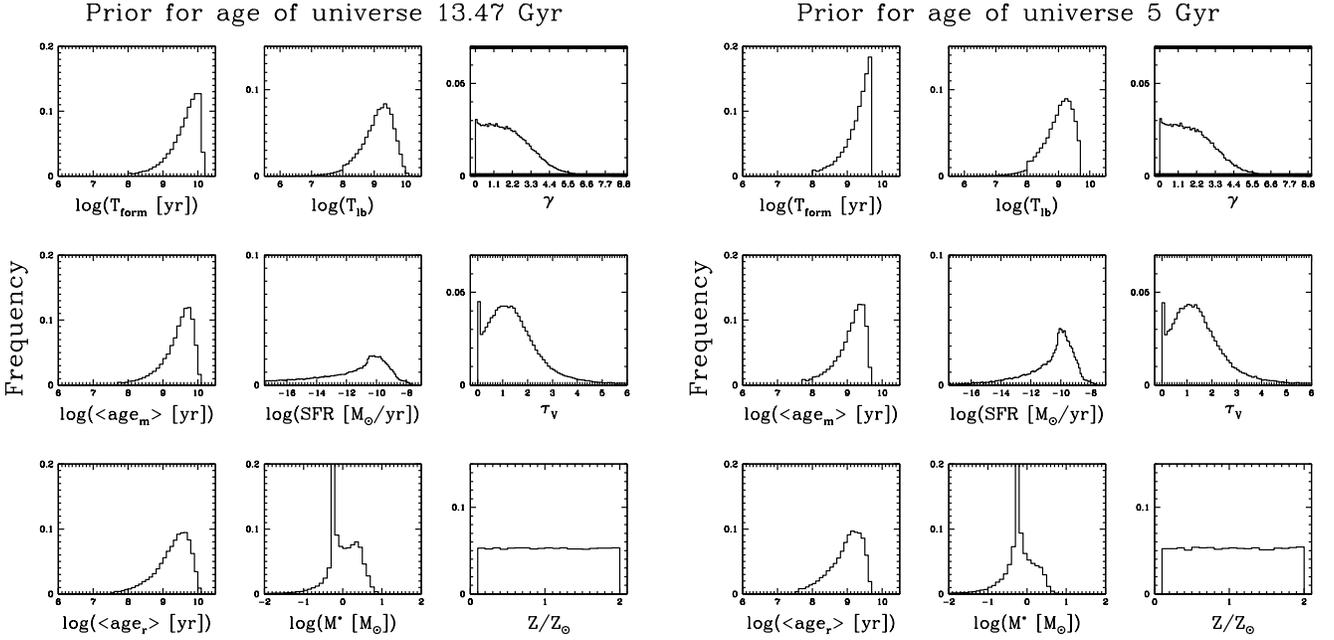

**Fig. 1.** Prior distribution of parameters in the library of stochastic star formation histories we are using. These are from bottom to top and from left to right: The mean r-band weighted age $\langle age_r \rangle$, the mean mass-weighted age $\langle age_m \rangle$, the age of the beginning of star formation $T_{form}$, the stellar mass $M^*$, the star formation rate in the mean over the last 100 Myr SFR, the age of the last burst of star formation $T_{lb}$, the metallicity in solar units Z, the attenuation parameter $\tau_V$, the parameter for the exponential fall-off of star formation $\gamma$ (see also Table 1 for definitions of the parameters). The prior changes with redshift, because we exclude model galaxies older than the universe from the library at each redshift. We therefore show the prior for redshifts zero (left) and 1.2 (right). Note that the parameters $M^*$ and SFR are subject to renormalization for each galaxy, depending on the luminosity of the objects.

It is noteworthy that the biases and uncertainties on the fitted parameters do not change significantly over our redshift range. There is some reduction in scatter, that is mostly due to the exclusion of models that are older than the age of the universe. Thus, the total possible age range of a galaxy at redshift 1.2 is less than half of what it is a z=0.1, which in turn reduces degeneracies between different SFHs.

In the remainder of this paper, we concentrate on the properties stellar mass $M^*$, star formation rate in the mean over the last 100 Myr SFR and attenuation parameter $\tau_V$. First, these allow the SFH of galaxies to be reconstructed and second the simulations just presented show that at least in principle useful measurements should be obtainable for these.

**Table 2.** Simulation results : mean offset, rms scatter

| Redshift | $M^*$ | SFR | $\langle age_r \rangle$ | $T_{form}$ | $\tau_V$ | Z |
|---|---|---|---|---|---|---|
| 0.1 | -0.02,0.3 | 0.03,0.5 | -0.1,0.2 | -0.3,0.3 | -0.1,0.8 | 0.06,0.4 |
| 1.2 | -0.06,0.1 | 0.05,0.3 | -0.07,0.2 | -0.1,0.2 | -0.08,0.7 | 0.06,0.4 |

Offset and scatter are given in log, but for the two parameters attenuation and metallicity.

## 3. Data

We now apply the SED-fitting method to a large, multi-wavelength dataset. Our extensive data were combined from a number of different surveys that provide multi-wavelength observations over the VVDS-F02 field of view as described below

### 3.1. Catalogues

Our primary starting point are the data from the F02 field of the VIMOS/VLT deep survey (VVDS), centred on RA = $2^h26'$ and DEC = $-04^d30^m$ over an area of $2 \times 2$ deg$^2$. This provides us with deep photometry from the CFHT/CFH12K camera in the B,V,R, and I bands with limiting magnitudes of 26.5, 26.2, 25.9, and 25.0, respectively (McCracken et al. 2003, Le Fèvre et al. 2004, http://cencosw.oamp.fr/). Additionally, deep NIR imaging down to limits of J≈ 21.50 and K≈ 20.75 is available for part of the sample (160 arcmin$^2$, Iovino et al. 2005). We also use the photometry from the Canada-France-Hawaii Telescope Legacy Survey (CFHTLS, http://www.cfht.hawaii.edu/Science/CFHTLS) field D1 over an area of 1 deg$^2$, in the version of the data release T0003. The data processing of the CFHTLS "deep fields" will be described in McCracken et al. (2008, in preparation). All objects in this sample have u∗, g′, r′, i′, z′ photometry down to limiting magnitudes of 26.5, 26.4, 26.1, 25.9, and 25.0, respectively.

We use the VVDS spectroscopic data acquired with the VIsible Multi-Object Spectrograph (VIMOS, Le Fèvre et al. 2003) installed at the ESO-VLT. These spectra give us spectroscopic redshifts for 8981 galaxies over ~0.5 deg$^2$ between $I_{AB}$ = 17.5 and $I_{AB}$ = 24 (Le Fèvre et al. 2005). Four redshift classes have been established to represent the quality of each spectroscopic redshift measurement, corresponding to confidence levels of ~55% (class 1), ~81% (class 2), ~97% (class 3), and ~99% (class 4) (Le Fèvre et al. 2005). The success in measuring redshifts depends somewhat on the spectral features that can be measured (see Ilbert et al. 2005 for a detailed analysis). For a more detailed description of the spectra themselves, see Section 3.2.

The F02 field was observed as part of the GALEX (Martin et al. 2005) Deep Imaging Survey in two channels (FUV 1530Å and NUV 2310 Å). We restrict the analysis to the central 1°diameter region of the GALEX field of view, which provides higher uniformity (Morrissey et al. 2005). In this region, number counts are 80% complete at 24.5 (AB system) in the FUV and NUV bands (Xu et al. 2005). The typical point-spread function (PSF) has a FWHM of 5″. Because of the high source density and relatively large PSF in these images, many sources cannot be deblended by source extraction algorithms. After a simple match between the CFHTLS optical catalogue and the GALEX photometry as derived from the pipeline, 45% of the UV objects have a double optical identification. In the case of the distant universe and for deep Galex pointings, simple associations with the closest or brightest optical counterpart are thus clearly susceptible to high uncertainties.

We therefore use a new approach to derive the photometry catalogues. Prior locations of potential UV emitters are determined from the optical sources in the field (objects with $u < 26.5$). Then each prior point source is convolved with the GALEX PSF. The flux in one pixel of the GALEX image is considered to be the sum of the contributions of all the priors and a mean background. This parametric model is optimal in the Poisson-noise regime. Individual fluxes are then assigned to each prior through a Maximum likelihood estimation. The method will be described in full detail in a forthcoming paper (Arnouts et al., in prep, see also Martin et al. 2007b). The reliability limit of the method has been estimated to be at a magnitude of 25.5. Those objects visible in the optical that are not detected in the NUV are assigned zero flux and a conservative error estimate of 0.23 $\mu$Jy.

The F02 field has also been observed as the XMM-LSS field of the Spitzer Wide-area InfraRed Extragalactic survey (SWIRE, Lonsdale et al. 2003), a large programme completed during the first year of the Spitzer Space Observatory. Infrared photometry in the IRAC 3.6, 4.5, 5.8 and 8.0 $\mu$m is thus available. We use the source catalogue as released by the SWIRE team (http://swire.ipac.caltech.edu/swire/swire.html), with detection limits at 5$\sigma$ of 5.0, 9.0, 43.0, 40.0 $\mu$Jy, respectively. Throughout the following we use the fluxes through the SWIRE aperture 2, i.e. with radius of 1″.9, as recommended by the data release paper.

Matching between all catalogues but the new UV catalogue is done inside the VVDS database CENCOS (LeBrun et al. 2008, in Prep., http://cencosw.oamp.fr/). Matching is performed according to spatial distance alone; the match radii have been optimized according to the spatial resolution of the data and are 0.5″between CFHTLS and VVDS photometry and 1″between VVDS and Swire photometry. The new NUV photometry being based on direct redistribution of flux according to optical priors from the CFHTLS data, matching is easily done by reference to the ID of the optical prior.

A combined multi-wavelength catalogue from the same field and based on the same data has already been used to measure the stellar mass density evolution based on a sample selected in the SWIRE 3.6$\mu$m band (Arnouts et al. 2007).

### 3.2. Available spectra and derived quantities

The VVDS observing programme was designed to obtain spectroscopic redshifts for galaxies between $I_{AB}$ = 17.5 and $I_{AB}$ = 24 (see Le Fèvre et al. 2005). The spectra have a resolution of R = 227 and typical signal-to-noise of 5 (see Figure 5 of Le Fèvre et al. 2005), thus optimizing the number of objects observed. The observed wavelength range goes from 5500 to 9500 Å. However, fringing and sky lines can considerably impair the spectra redwards of 8000 Å.

Albeit not comparable in quality to the spectra of the Sloan Digital Sky Survey (SDSS), such spectra from a survey of much higher redshift galaxies also contain information that can be extracted. First, for galaxies with emission lines, one can reli-tracted emission line fluxes. For this paper we are in particular interested in the H$\alpha$ line (for objects between redshift of 0.0 and 0.44) and the [OII] line (for objects between redshift of 0.48 and 1.2). On average, the errors on the detected lines are 20% and 30% for [OII] and H$\alpha$, respectively. We here only use these emission line measurements as a check on the reliability of the SED fitting. A detailed study of the physical properties of VVDS galaxies from their emission lines will be undertaken in Vergani et al. (2008, in prep.).

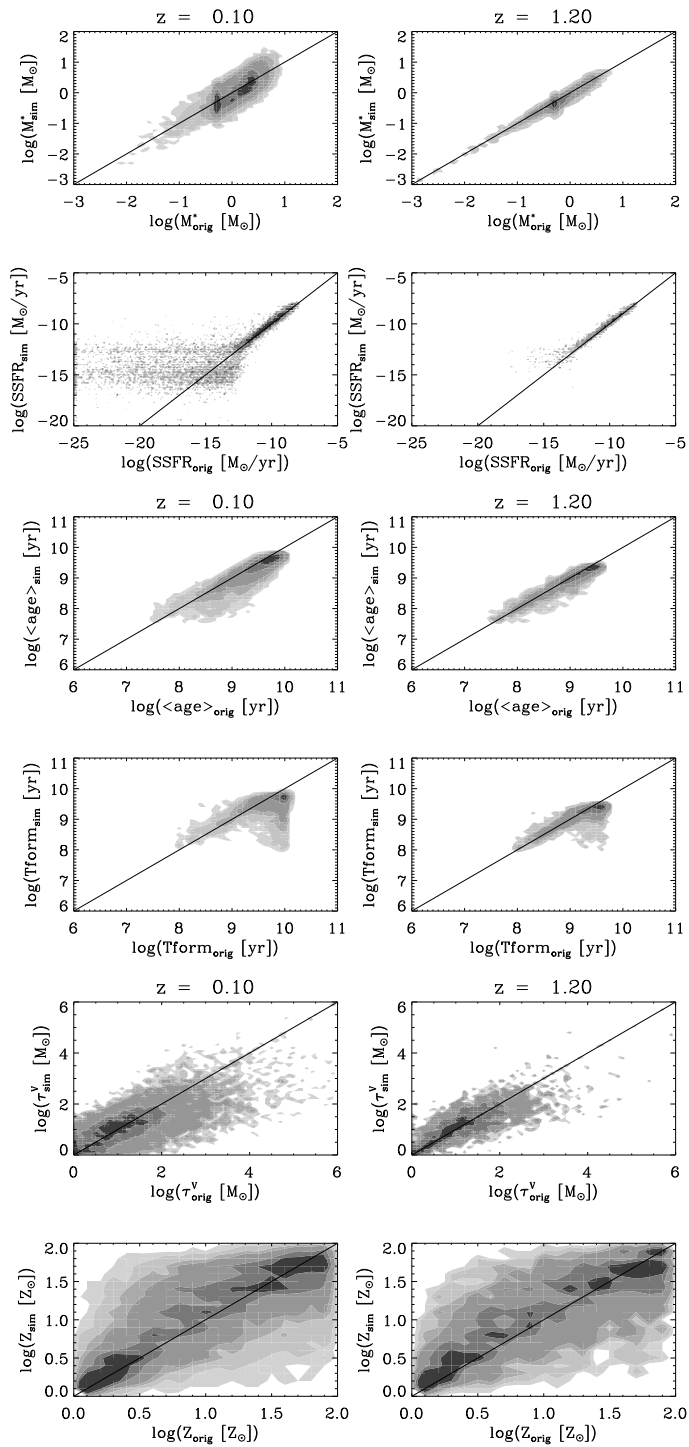

**Fig. 2.** Results of fits to noised "pseudo-galaxies" produced from the library of star formation histories. The plots show the fitted parameters vs. those in the original library over a range in redshift. Note that stellar mass M* and mean age $\langle age_r \rangle$ are well determined, as well as the SSFR above a minimal threshold. On the other hand the formation age and the metallicity are not well determined, *even in the case of a perfect model.*

An important issue when comparing spectral and photometric information is the effect of a mismatch between the aperture used to extract the photometry and the spectra. However, this effect is less severe in the VVDS than in lower redshift surveys for two reasons. 1) The VVDS is based on long-slit spectra and extraction apertures are optimized to contain all the visible flux along the slit. For this reason alone, aperture biases are much smaller than for fiber-based spectral extraction. 2) Most of the galaxies are small on the sky due to their distance.

Correction factors (in the following called "normalization") for the VVDS were derived by synthesizing photometry from the spectra using the instrumental response functions of the instrument. These aperture normalizations are released to the public together with the spectra. In the VVDS R band we find a median normalization factor over the full sample of $F_R/F_{spec} = 1.3$ with an rms of 0.5. This normalization factor is almost independent of the redshift and size of the object on the sky. The largest contributors to the scatter in this factor are the different observing conditions, as the survey spectra were assembled over an extended period of time. The physical properties of a galaxy change with radius. Therefore the stellar populations probed by a photometry that includes all light from a galaxy and those populations probed by a spectrum that was taken over only a fraction of the light from the same galaxy can in principle be biased. As we find no dependence of the normalization factor on either size of the galaxy or its colour, we made no further attempt to correct for such issues.

### 3.3. Sample selection and catalogue cross-match

We define two subsamples of objects for different purposes. Our first sample, which we call the "photometric sample" is designed to be as complete as possible, providing a comprehensive view of galaxy properties over the age of the universe. The second sample, which we call the "spectroscopic sample", is designed to provide us with secure spectroscopic data, in order to test the validity of our approach.

The "photometric sample" comprises all objects that have been detected at $18 < I_{AB} < 25$ in the VVDS data over the area covered by the F02 field. At this flux limit, the VVDS and CFHTLS data are complete. For all of these objects we have spectroscopically calibrated photometric redshifts from Ilbert et al. (2006) with an accuracy of $\sigma_{\Delta z/(1+z)} \approx 0.03$ for objects brighter than $i'_{AB} = 24$. We cut this sample at a redshift of 1.2. The reddest band regularly available for all galaxies, the CFHTLS z' band, corresponds to roughly 4000 Å at this redshift. Redwards of this wavelength, the SED is dominated by stars from the main-sequence turnoff, thus allowing the mean age of a galaxy to be effectively probed. Bluewards of 4000 Å, the SED starts to be dominated by stars that emit strongly in the UV, i.e. O and B stars, which are in turn only associated with the recent star formation. At redshifts higher than 1.2 we would thus enter a totally different regime, where the SED is constrained by restframe UV information only (see e.g. Law et al. 2007) and ably identify Seyfert 1 nuclei (Gavignaud et al. 2006), which are excluded from the present sample. Second, the VVDS team has adapted the `platefit` programme designed for the SDSS (Tremonti et al. 2004) to the lower resolution VVDS spectra (Lamareille et al. 2008). This software yields continuum sub-where the SED does not provide a reliable measure of the mean age or stellar mass of a galaxy. If NIR photometric data are available for an entire sample, there is no other obstacle to prevent the extension of the present methods to higher redshifts.

Our photometric sample contains 84073 galaxies. All of them have B,V,R,I,u∗, g′,r′, i′,z′ photometry; 39% are detected in the NUV, 5% in $J$, 17% in $K$, 29% in the IRAC 3.6 and 4.5 $\mu$m bands simultaneously. Eight hundred twenty-five objects are detected in all bands from NUV to 4.5 $\mu$m. Our total photometric catalogue is built on the available photometry in the observed optical wavelength. The total area covered by the combined photometric catalogue is thus the cross-section between the CFHTLS D1 photometry and the VVDS photometry, i.e. 0.89 deg$^2$.

For the "spectroscopic sample" our goal is to assess how feasible it is to recover physical parameters for galaxies from population synthesis models. Starting from the photometric sample, we therefore restrict ourselves to objects with spectra and with VVDS redshift classes with a confidence level greater than or equal to 81% (classes 2,3 and 4). This is a compromise between our wishes to avoid overly biasing ourselves to galaxies with spectral features that make redshift measurements more secure and to avoid fitting galaxies under completely wrong assumptions.

After matching and excluding galaxies with redshifts higher than 1.2, we have a total sample of 5753 galaxies in the spectroscopic sample. Out of these, all have by definition observed optical photometry in the B,V,R,I,u∗,g′,r′, i′,z′ bands, as well as spectra in the observed optical wavelength region. Additionally, 67% are detected in the NUV, while an additional 17% have upper flux limits. Twelve percent and 48% are detected in $J$ and $K$, respectively. Finally, while 45% are detected in the IRAC 3.6 and 4.5 $\mu$m simultaneously, only 3% are also detected in the 8.0 $\mu$m band due to the shallower detection limit. Out of our total 5753 objects, 212 are detected in all bands from the NUV to 4.5 $\mu$m. Additionally, as described in Section 3.2, information on emission lines properties is available for all objects from their spectra. For reference, the redshift distributions of our two samples are shown in Figure 3.

## 4. Data meet library

We now describe the results of applying the method described in Section 2.1 to the data as described in Section 3 using the model library as described in Section 2.1.

### 4.1. General applicability of the models to our data

We first checked, whether the observed and the model galaxies are similar in their photometric properties. We find that, in general, the range of observed and modelled colours do cover the same loci in parameter space. This confirms that the stellar population model is a good representation of real galaxies over all the redshifts probed in this paper. Furthermore, as the density distribution of observed and model galaxies in each region of parameter space is similar, we also have some reality check on the assumptions that went into our prior distribution, as presented in Section 2.1.

In Figure 4 we show one of our checks. As can be seen from the figure, the loci of data (photometric sample) and model library colours move with redshift. This is to be expected, as the rest frame wavelength region observed through a particular filter depends on redshift. For most redshifts, the locus of data and model galaxies overlap. However, for one specific redshift, i.e. for a redshift of 0.7, the contours indicating the locus of the model library move differently from the locus of the data - for the colour combination we show here. A very similar effect is also seen in data accumulated for use in the DEEP2 survey (S. Salim, private communication). At redshift 0.7, the models are bluer than the data in $g - r$, and redder than the data in $r - i$. This shift of approximately 0.5 mag in the r-band is mostly observed for blue (young) galaxies. It can be explained if there is a restframe wavelength region in which the flux is underpredicted by the stellar population model.

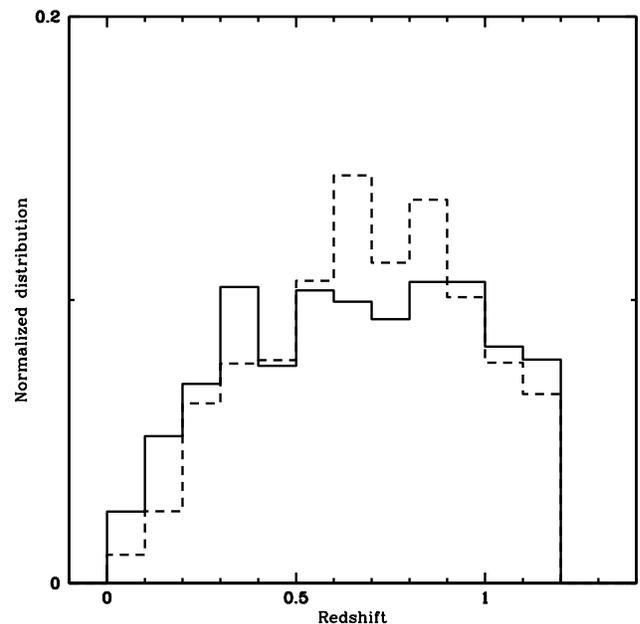

**Fig. 3.** Redshift distribution of the 5753 galaxies in the "spectroscopic sample" (dashed line), i.e. with secure spectroscopic redshifts, and of the 84073 galaxies in the "photometric sample" (full line), i.e. with high quality photometric redshifts.

The observed $r$-band ranges from 5600 to 6900 Å, approximately. This samples a rest frame wavelength region between 3300 and 4050 Å at a redshift of 0.7. It has been shown (Wild et al. 2007) that the BC03 models overpredict the strength of the Balmer break for young galaxies. As there has been little change in this wavelength region in the preliminary version of CB07 we are using, we assume that this is just another manifestation of the same model limitation. For each object we therefore exclude the filters that draw more than 5% of the total light from the so-defined restframe wavelength region from any later analysis. We have verified that excluding those photometric bands from the fit avoids a bias in the results towards younger ages and higher SFRs.

We also exclude the bands measuring flux at wavelengths longer than 4$\mu$m due to possible contamination from hot dust or PAH emission. The bands used for the fitting at each redshift are summarised in Table 3.

### 4.2. Quality of fit

To apply the Bayesian method, we calculated $\chi^2$ for every combination of object and model galaxy in our two samples and in the library. As $\chi^2$ is a measure of the "goodness-of-fit", we can also use the minimum $\chi^2$ we find for each object to gauge the applicability of our model to our data.

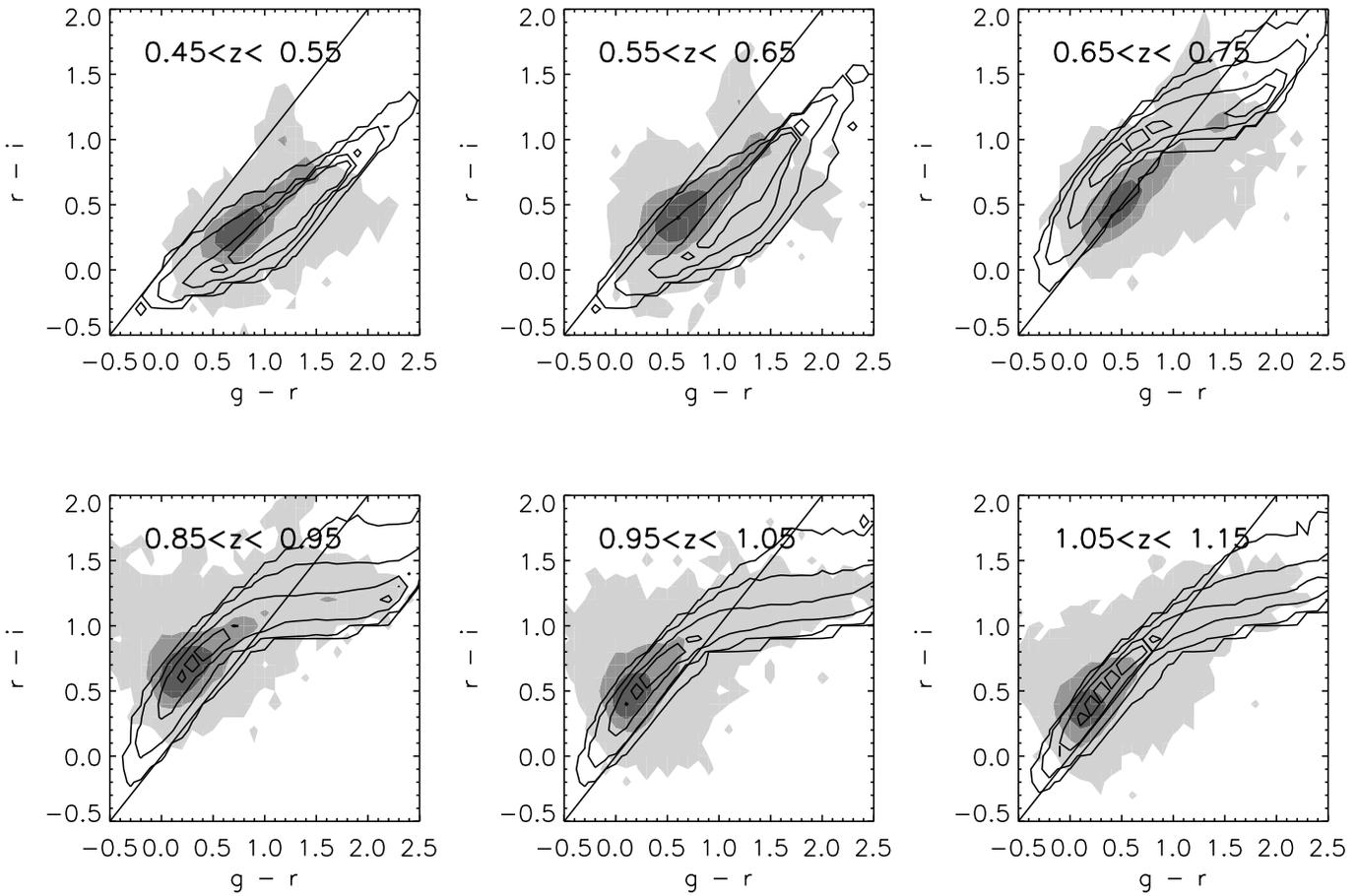

**Fig. 4.** Overlap between the loci in colour-colour space of data and models. The observed data are in grey shading, while model galaxies are shown as solid contours. In general the overlap between the loci covered by the observation and the model library is good. However, this plot suggests that the spectral region between restframe wavelengths 3300 to 4050 Å is not correctly reproduced by this preliminary version of the CB07 stellar population model. We have verified that similar problems do not occur at other wavelengths.

**Table 3.** Bands used in the fitting

| Redshift | NUV | u* | B | g' | V | r' | R | i' | I | z' | J | K | 3.6μm | 4.5μm | 5.8μm | 8.0μm |
|---|---|---|---|---|---|---|---|---|---|---|---|---|---|---|---|---|
| 0.1 | x | - | - | - | x | x | x | x | x | x | x | x | x | - | - | - |
| 0.2 | x | - | - | - | x | x | x | x | x | x | x | x | x | - | - | - |
| 0.3 | x | x | - | - | - | x | x | x | x | x | x | x | x | - | - | - |
| 0.4 | x | x | - | - | - | x | x | x | x | x | x | x | x | x | - | - |
| 0.5 | x | x | x | - | - | - | - | x | x | x | x | x | x | x | - | - |
| 0.6 | x | x | x | - | - | - | - | x | x | x | x | x | x | x | - | - |
| 0.7 | x | x | x | x | - | - | - | x | x | x | x | x | x | x | x | - |
| 0.8 | x | x | x | x | x | - | - | - | - | x | x | x | x | x | x | - |
| 0.9 | x | x | x | x | x | - | - | - | - | x | x | x | x | x | x | - |
| 1.0 | x | x | x | x | x | - | - | - | - | x | x | x | x | x | x | - |
| 1.1 | x | x | x | x | x | x | - | - | - | - | x | x | x | x | x | - |
| 1.2 | x | x | x | x | x | x | x | - | - | - | x | x | x | x | x | - |

From the fit results on the photometric sample, we find that, in general, the model can be said to reproduce the available data fairly. The median reduced $\chi_\nu^2$ is 1.35, while the mean is 4.4, excluding outliers that have $\chi_\nu^2 > 100$ (see Figure 5). We explicitly note here that the distribution in $\chi_\nu^2$ would be somewhat tighter if we had multiplied all error values by a constant factor as done e.g. in Ilbert et al. (2006) and Iovino et al. (2005). Out of our photometric sample of 84073 galaxies, 0.4%, 2.0% and 7% have $\chi_\nu^2$ larger than 500, 100 and 20, respectively. A browse through such poorly fit objects shows that outliers can be caused by a combination of two effects. 1) While for such a large survey, uniform procedures of data reduction and matching are clearly necessary, there are always objects where such standard procedures fail (e.g. extended objects, catalogue mismatches, blended targets, Sey2 AGN, and unusual targets) 2) Due to the high quality and wide wavelength coverage of the data we use, the model reaches its limits.

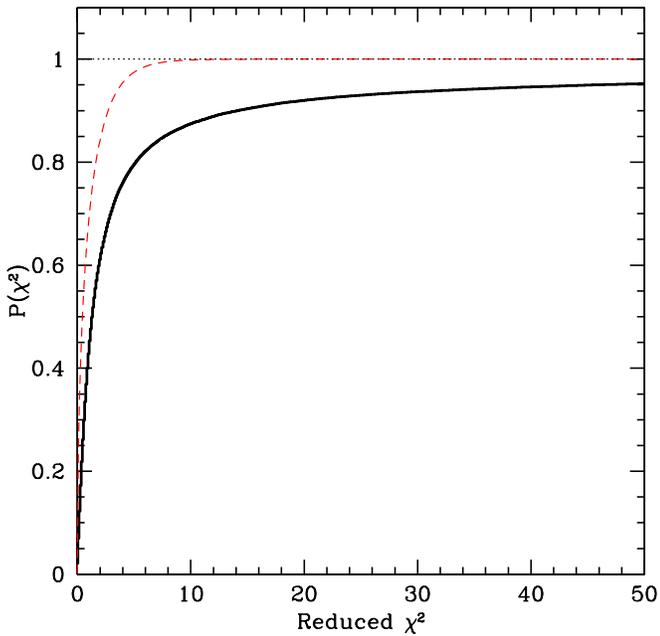

**Fig. 5.** Cumulative distribution of reduced $\chi^2$ (thick solid line) as compared to the expectation for 1 degree of freedom (thin dashed line). The lines for more degrees of freedom would be even steeper.

Example fits to the photometry are shown in Figure 6 for two objects that are detected in all bands from the NUV to 8 $\mu$m.

### 4.3. Stellar masses

Stellar masses for the VVDS spectroscopic sample have been derived by Pozzetti et al. (2007, P07) by an SED-fitting technique essentially identical to ours. In particular these authors have tested the influence of the prior distribution of SFHs on the derived stellar masses. It is of interest to verify that the stellar masses derived in the present paper are compatible with those of P07. To that end we determined the physical properties of the galaxies not only with the CB07 models, but also with the BC03 ones, as these were used in P07. Otherwise our setup was left unchanged. We used the masses from P07 that were determined from models allowing for secondary bursts. We find that our results agree with those of P07 to within 10%, as expected.

Without independent kinematic data, there is no direct way to assess how correct in an absolute sense the stellar masses are that we determine. Di Serego Alighieri et al. (2005) find, however, that the Chabrier (2003b) IMF we use results in masses consistent with the dynamical ones. An in-depth comparison of the effects of using different stellar population models (BC03 and M05) on the determined stellar masses can be found in Van der Wel et al. (2006). These authors use dynamically determined masses as a benchmark. They find that BC03 and M05 models bracket the mass value determined dynamically, with offsets of the order of a factor 2. Also, the use of complex SFHs, as done in the models used here, can significantly improve the correctness of the resulting stellar masses. As the contribution of the TP-AGB stars to the total infrared light in the CB07 model is intermediate between the BC03 and the M05 models, and as we are allowing for secondary bursts in the SFH, we expect our stellar masses to be robust and correct in an absolute sense.

The robustness of stellar mass measurements in a relative sense is more easily assessed. First, photometric data in the $J$ and $K$ bands or at 3.5 $\mu$m (collectively referred to as near-infrared, NIR) is available only for a part of our sample. A thorough discussion of the uncertainties in stellar mass arising from limitations in data coverage and from the use of different priors can be found in P07. From our own sample we confirm the finding of P07 that exclusion of the NIR data leads to an overestimation of the stellar mass at the high-mass end of the sample. When correlating this bias in the estimation of $M^*$ with other physical parameters, we find that the strongest correlation is with SSFR. The mean ratios of masses determined without NIR data to the masses derived with NIR data are 2.8, 1.50, 1.0 for bins of log(SSFR) of [-16,-13],[-13,-10],[-10,-8], respectively. These correction factors are independent of redshift, within the uncertainties. Similarly to P07, we correct for this effect by dividing the masses of objects that have not been observed in the NIR by these ratios. This approach relies on the assumption that the objects that have not been observed in the NIR have the same properties as those that have been observed.

Finally, we mention that a direct comparison of the new CB07 models to the standard BC03 models shows a rather modest offset in stellar mass. This offset changes with the measured SSFR of the galaxy, in the sense that the masses determined using BC03 for objects with an intermediate SSFR ( -12 < log(SSFR) < -9) are in the mean more massive by 50% than those determined using CB07. These are the objects where the TP-AGB stars are most likely to contribute significantly to the light in the NIR bands. Objects with lower or higher SSFRs do not show this offset. Our results agree with the findings of Maraston et al. (2005, 2006), who for the first time showed the significant effect of the TP-AGB phase of stellar evolution on the derived stellar masses.

### 4.4. The relation between attenuation and star formation rate

From the fit results we can show the relation between SFR and $\tau_V$. Figure 7 presents $\tau_V$ as a function of star formation rate in 4 different redshift bins. The two quantities show a broad relation in all the 4 redshift bins. In our two lowest redshift bins we sample a comparatively small volume, thus missing rare, high star formation rate objects (see Ilbert et al. 2005). On the other hand, we are sampling brighter galaxies at higher redshift. That we observe different populations of galaxies at different cosmic epochs (Tresse et al. 2007) affects the mean measured attenuation. On the other hand, the scatter remains large at every epoch, translating the diversity of galaxy properties.

We also plot, for reference, the relation between SFR and $\tau_V$ as inferred from the dependence of the Balmer decrement on star formation rate found by Sullivan et al. (2001) in a sample of nearby galaxies. These authors also explicitly note the large scatter around the mean relation. For simplicity, we converted H$\alpha$/H$\beta$ into E(B-V) using the formulae described in Reynolds et al. (1997), which relies on the Osterbrock (1989) interstellar extinction curve. The necessary conversion does not depend sensitively on the adopted dust prescription, because of the small wavelength leverage involved. To convert E(B-V) to $\tau_V$ we additionally assume $R_V = 3.1$ (see e.g. Cardelli, 1989). Although our measurement of the attenuation concerns the stellar continuum, and the Sullivan et al. (2001) measurement is based on emission lines, the general agreement shown in Figure 7 can be considered to be quite good, in particular in light of the intrinsic uncertainty on the attenuation parameter as described in Section 2.2. We additionally plot several literature relations in the lowest redshift bin: from Hopkins et al. (2001, their Eq. 3), based on the Balmer decrement of local starburst galaxies (dot-dashed); from Buat et al. (2007, dashed line, their Figure 7), based on the ratio of the total infrared luminosity over the UV luminosity of a local sample of galaxies bright in the far infrared and on relations given in Hirashita et al. (2003); from Choi et al. (2006, dotted line, their Section 5.2.1), who derive their attenuation estimate from a comparison of the optical SFR to the FIR SFR, again based on a sample of starburst galaxies. Finally, in the uppermost redshift bin, we also show the relation we determine from Burgarella et al. (2007, dashed line, their Figure 6), again using Hirashita et al. (2003). While we note that we tend to measure lower SFRs at the same attenuation values, this is most probably due to the literature relations having been derived for samples with a restricted range in SFR. Indeed, all these last relations are based on samples, where strongly star-forming galaxies are well represented by selection.

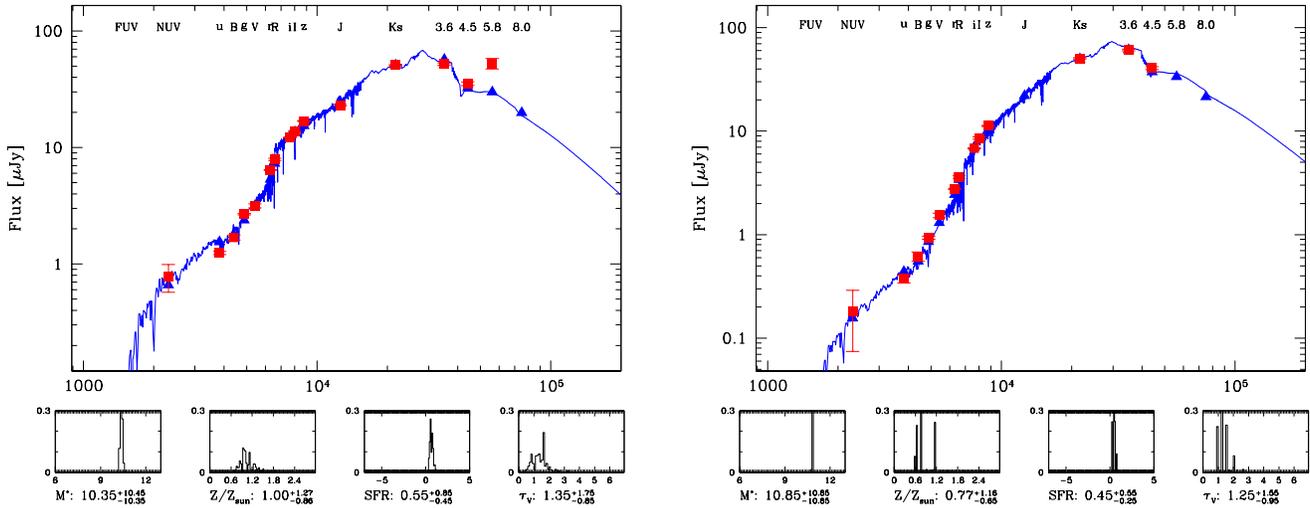

**Fig. 6.** Two example SEDs that were observed and detected in all bands from the NUV to the 4.5 μm band of Spitzer/IRAC (filled squares with errorbars). The best-fit model is shown in two ways: the spectrum as a solid, thin, line and the synthesized photometry as triangles. The population synthesis model is in general able to reproduce the observed SEDs. Nevertheless these examples also highlight two problems we generally encounter and have to account for: on the left, it can be seen that the 5.8 μm band of Spitzer/IRAC deviates significantly from the Rayleigh-Jeans type falloff we would expect. This can probably be attributed to PAH emission lines or hot dust. On the right, close inspection shows that the restframe wavelength region around 3500 Å in the model is affected by the problem discussed in Section 4.1. The panels on the bottom show the PDF for four different parameters, as well as the median and the upper and lower limits to the confidence intervall. Note also that the (normalized) probability distribution functions are more narrow for the parameters stellar mass $M^*$) and star formation rate (SFR) and broader, or even with several maxima, for attenuation ($\tau_V$) and metallicity (Z).

Our plot of SFR vs. $\tau_V$ is at least qualitatively similar to Figure 1 in Calzetti (2007). Finally, Salim et al. (2007) also find that on the whole the attenuation as measured from SED fitting agrees reasonably well with the one determined from accurate modelling of the emission lines, although the SED fitting value is less tightly constrained.

In the median over the sample, the attenuation at the wavelength of Hα is a factor of 3. This is consistent with the mean extinction of the Hα line for local samples of galaxies, which is very roughly 1 mag (Kennicutt 1998), i.e. a factor of 2.5.

We conclude that, while the extinction measurement on a galaxy by galaxy basis is uncertain, it can be considered to be realistic in the average over the photometric sample. This last statement is consistent with the results from simulations shown in Section 2.1.

### 4.5. Comparison to spectral measurements

We now use the spectroscopic subsample to check whether we can trust the SFRs as derived from the photometry. Here we are interested in the SFRs we can derive for the galaxies in the spectroscopic sample from the Hα and OII emission lines. The spectra themselves contain more information, which has been used in Vergani et al. (2007) and Franzetti et al. (2007). We emphasise that we redetermine the physical properties of the objects using the redshifts as determined directly from the spectra for this comparison. Because we use identical redshifts, our comparison bears on the *physical* relation between both methods to determine SFRs.

SFRs for the spectroscopic sample have been determined in Vergani et al. (2008, in prep.). In order to obtain the SFRs from the OII line, these authors use the calibration presented by Moustakas et al. (2006). This has the advantage of also utilising the information contained in the B-band absolute magnitude, thus accounting, at least statistically, for the effects of dust atten-uation and varying metallicity on the derived SFR. Nevertheless, using this calibration at higher redshifts assumes that the correlations between B-band luminosity and attenuation/metallicity do not change significantly with redshift. The H$\alpha$ SFRs were derived using the standard Kennicutt (1998) formula.

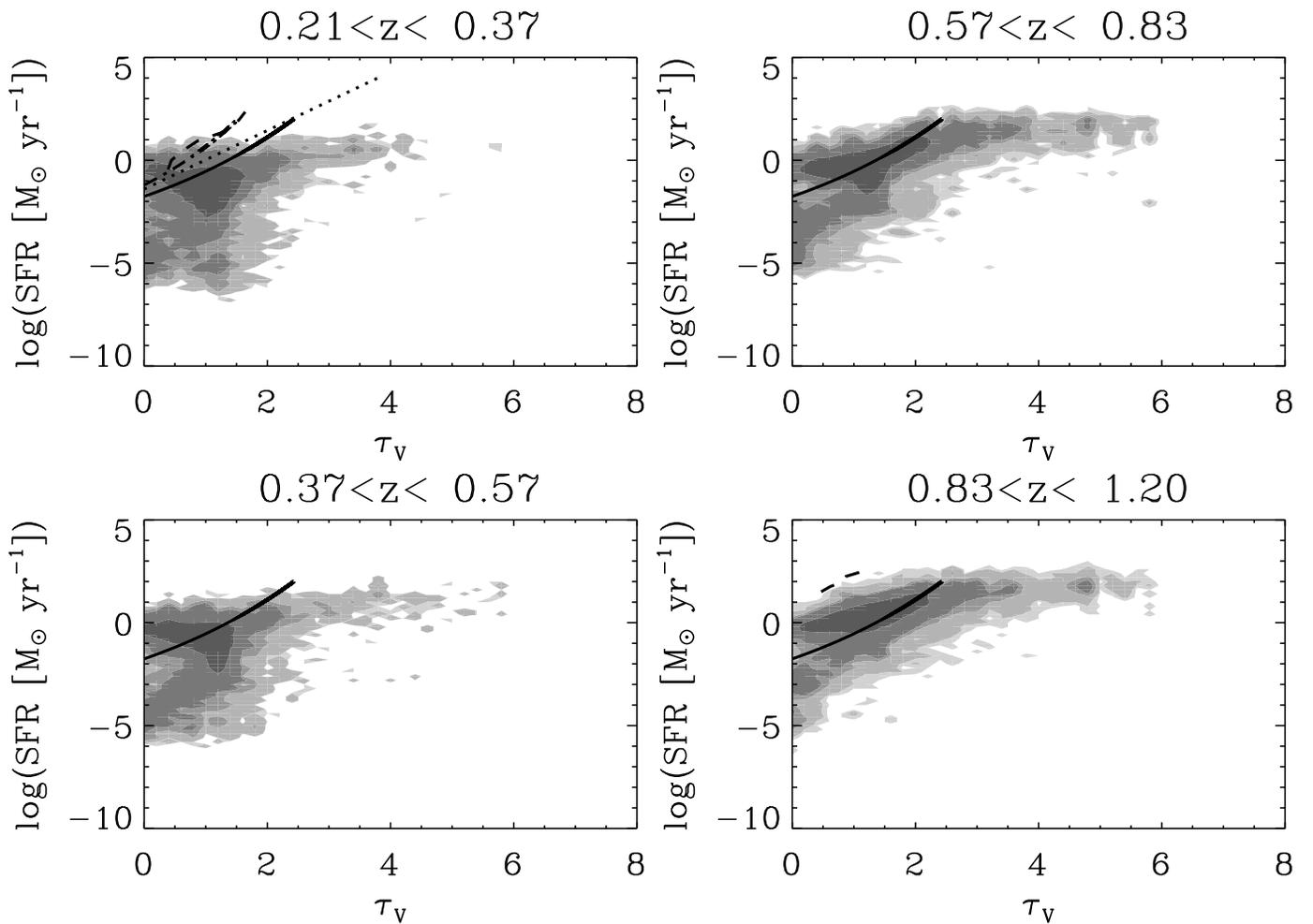

**Fig. 7.** Relation between SFR and attenuation $\tau_V$ as a function of redshift. The solid line is a local universe comparison from Sullivan et al. (2001). The dotted and dashed lines present more literature relations, as described in the text. Note that the two lowest redshift bins sample a small volume and therefore galaxies with stellar masses similar to those of the Sullivan et al. (2001) sample are under-represented.

The correlation between the SFR determined in this way and the SFR determined from the SED fit is shown in Figure 8 for both emission lines. The total number of objects in these plots (grey plus signs) is 2776 (for OII) and 679 (for H$\alpha$), and 52% (OII) and 45% (H$\alpha$) of those are considered to be of high quality (black plus signs), as determined in Vergani et al. (2008). If we define catastrophic failures as objects with $\log(\text{SFR}_{EL}/\text{SFR}_{SED}) < -1.5$, the percentage of such catastrophic failures for the total sample is 11% and 10% for OII and H$\alpha$, respectively. Excluding catastrophic failures, we obtain: the median of $\log(\text{SFR}_{H\alpha}/\text{SFR}_{SED})$ is 0.07 and the scatter in log is 0.71. The median of $\log(\text{SFR}_{OII}/\text{SFR}_{SED})$ is -0.02 and the scatter in log is 0.62. Thus, both determinations appear to be consistent.

Nevertheless, two important caveats need to be mentioned that should in principle hamper the direct comparison of the two SFR indicators.

First, star formation is episodic (e.g. Glazebrook et al. 1999). However, the UV light of stellar populations fades more slowly than the emission line flux, thus the UV gives the average star formation rate over a longer period of time. In our case the SFR as derived from the photometry is an average over the last 100 Myr, while emission lines tend to disappear soon after a burst on a timescale of 10-20 Myr. Thus, the typical duration of a star formation burst will influence how well these two measures of SFR can be intercompared. We can empirically estimate the size of this effect using the following recipe. In the interest of higher and less biased statistics, we go back to a sample including ALL spectra that were measured in the F02 here. There are in total 8981 spectra in the F02. Out of these, 3758 have successfully measured emission lines (either [OII] or H$\alpha$); yet, according to the photometric fit, 7040 have an SFR above $\log(\text{SFR}) = -0.2$[4]. At face value, this factor of $\approx 2$ difference between the number of photometrically star-forming galaxies and those with detectable emission lines could be accounted for if the typical duration of a burst is half the time over which the SFR is averaged, i.e. 50 Myr. Then, the SFRs as measured from the emission lines would need to be scaled down by a factor 0.5 to yield the same average SFR as the one used in the SED fit. The true correction is probably smaller than indicated by this number, as there are a number of other reasons, why objects might not have a measured emission line. Among these reasons are faint emission lines, extraction problems, misalignment of the slit and more.

---

[4] This is a crude estimate for the SFR equivalent to the detection limit for the [OII] emission line. Here we are only interested in a rough estimate of the size of the potential correction. A more detailed study, which is beyond the scope of the present paper, would have to take into account how this limit varies with redshift and would need to make a more complex model, in which bursts can have a range of durations.

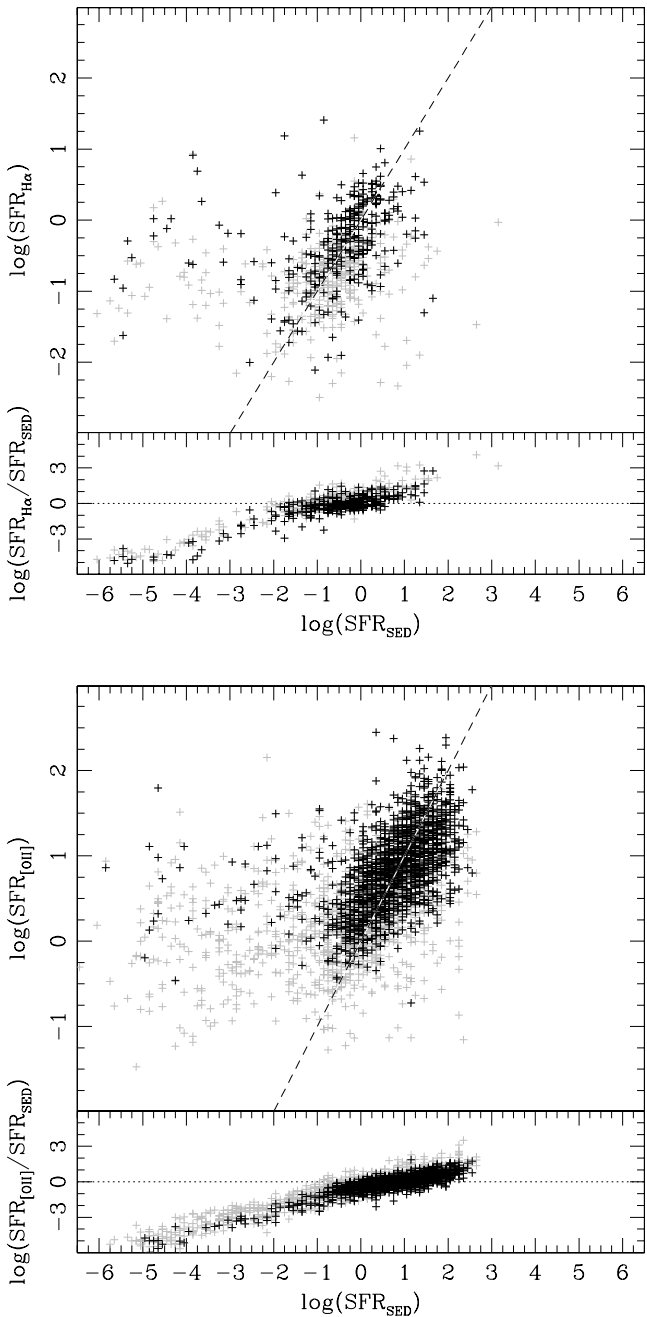

**Fig. 8.** Relation between SFRs as determined from the photometry and as determined from the Hα (upper panel) and [OII] (lower panel) emission lines. Black symbols show those objects for which the emission line has been detected with good confidence, while light grey points have upper limits or marginal detections. The dashed line shows the one-to-one correlation.

Objects that have only upper limits on the strength of the emission line are shown as grey symbols in Figure 8. The SFR values were obtained by converting the flux value corresponding to the $1\sigma$ detection limit into a SFR. This verifies that their photometrically determined SFR is compatible with emission below our detection threshold for most objects. In the same context, the scatter in Figure 8 is dominated by the catastrophic failures that can be seen as a spur of objects for which the $SFR_{SED}$ is much lower than $SFR_{OII}$. It is possible that these catastrophic failures represent objects for which star formation has just begun and where most of the newly formed stars are heavily obscured, thus having no impact on the observed SED (compare Da Cunha et al. 2008).

Second, while the sample has been cleaned for Sey1 AGN, the selection of type 2 AGN from the VVDS spectra is still in progress and we have not eliminated these objects from our sample. Locally the fraction of light in the [OII] and Hα line emitted by AGN is 30% and 25% respectively. These values have been computed from the catalogue of galaxies analysed in Brinchman et al. (2004). These numbers give a very rough estimate of the possible contribution that Sey2 AGN would have to the total emission in the respective emission lines.

If we applied the derived corrections, the median offset in $\log(SFR_{OII}/SFR_{SED})$ in dex would be changed from -0.02 to -0.47. We conclude that overall the SFRs as derived from the two completely independent tracers agree with each other inside the, considerable, measurement uncertainties. This conclusion has also been reached in studies of the local universe (Salim et al. 2007, Treyer et al. 2007).

While not presented here, we point forward to Wild et al. (in prep.), in which a direct comparison between the continuum properties of the spectra and the SED fitting properties is undertaken for a subclass of galaxies, the post-starburst galaxies. There we find again that the SED properties are entirely consistent with the completely independent spectral measurement.

## 5. Following mass and star formation rate over cosmic time

In the last section we have derived M* and SFR for a sample of 84073 galaxies in a redshift interval between 0 and 1.2. This is the first catalogue to hold such information for such many galaxies and for galaxies with star formation rates too low to have detectable emission lines or IR emission. We now proceed to take a first look at the information that this catalogue provides on the evolution of star formation and stellar mass over redshift. More in-depth studies of these issues will be undertaken in the forthcoming papers by Lamareille et al. (2008, in prep.) and Vergani et al. (2008, in prep.). From now on we split our sample into four redshift bins that are approximately equally spaced in cosmic time, i.e. 1.5 Gyr wide. The respective lower, and upper redshifts are: [0.21,0.37],[0.37,0.57], [0.57,0.83],[0.83,1.20].

### 5.1. Incompleteness

While the completeness in a particular photometric band can be judged empirically, this is not possible for the completeness in derived parameters, such as M* and SFR. We therefore use an alternative approach to derive an estimate of the completeness of our photometric sample. We first note that our ability to detect a specific galaxy will depend on its stellar mass, on the mass-to-light ratio, and on the shape of the SED (see e.g. Ilbert et al. 2004 for an in-depth discussion on the effect of this bias on the determination of luminosity functions). The two last parameters are determined by the ratio of recent to total star formation history of the galaxy, which is conveniently parameterized in terms of the specific star formation rate SSFR = SFR / M*. We assumed from the beginning that our library of model galaxies is representative of the population of galaxies at the redshifts surveyed in this paper. Figure 9 shows the percentage of model galaxies brighter than $I_{AB} = 25$ as a function of M*, SSFR and redshift. It need not be emphasised that we cannot correct for incompleteness using the completeness array as this would imply the assumption that we not only sample the parameter space of real galaxies, but actually know the relative densities of the galaxies in this parameter space. For comparison we also plot the 80% completeness limit derived empirically by Meneux et al. (2007) for the VVDS data, however for a magnitude limit at $I_{AB} = 24$. This mass limit is compatible with Figure 9, if one takes into account that the real distribution of galaxies in SSFR is non-uniform, with most observed galaxies lying in the two uppermost dex of the plot. Note also that the spacing between the iso-completeness lines in the plot depends strongly on the *assumed* prior and cannot be directly compared to the real data.

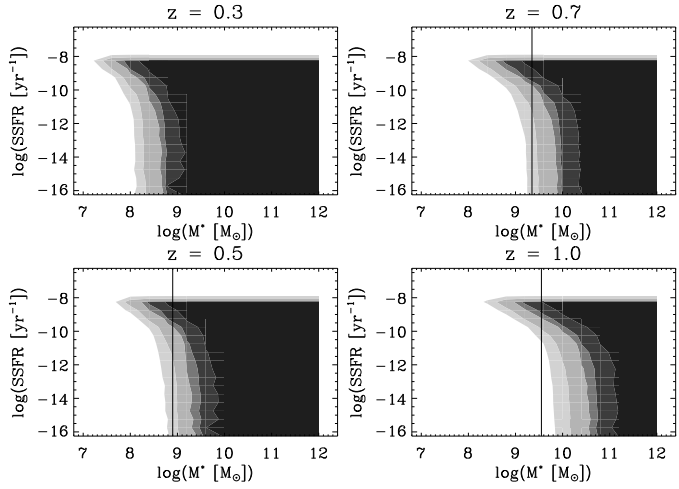

**Fig. 9.** Percentage of model galaxies in our library that have $I_{AB} < 25$ as a function of SSFR and stellar mass. White means no galaxy is detected, while black means all galaxies will be detected. Isocontours are at 30%, 60%, 90%, 95% and 99%. The solid vertical line shows the 80% completeness limit derived empirically by Meneux et al. (2007) for $I_{AB} > 24.0$.

For later use we now define the completeness limit in M* to be where we are 50% complete in the model catalogue, marginalized over all SSFRs. According to this criterion, the completeness limits are at $\log(M_\odot) = 9.2, 9.6, 10.0$ for $z = 0.5, 0.7, 1.0$, respectively. This would correspond to the redshift bins [0.37,0.57], [0.57,0.83], [0.83,1.2], again respectively.

### 5.2. The number density of galaxies as a function of stellar mass and specific star formation rate

In Figure 10 we show our measurement of the space density $\Xi$ of galaxies as a function of M*, SSFR and redshift for a large sample of galaxies and for all star formation rates. In the following we call $\Xi$ the "Mass-SFR function" analoguous to the widely used luminosity function. The space density of objects was derived using the classical $V_{max}$ method (Schmidt, 1968; Felten, 1976) for luminosity functions. This weights galaxies according to the fraction of the survey volume in which they could have been observed, given their brightness and the shape of their SED. In our case, we simply scale the best-fit model to the observed galaxy magnitude and shift it in redshift to determine the total redshift path in which the galaxy would be seen, given the survey magnitude limits. The comoving volume sampled by the survey between the lower and upper redshift limits of the bin was computed following Carroll et al. (1992). The area covered on the sky is $A_{survey} = 0.89 \deg^2$.

The objects with measured SSFR below $10^{-14}$ yr$^{-1}$ were set to $10^{-14}$ yr$^{-1}$. As shown in Section 2.2, the lower limit where we can trust the measurement of the SSFR is at roughly $10^{-12}$ yr$^{-1}$. A dashed line shows this limit. The apparent "sequence" at $10^{-14}$ yr$^{-1}$ is therefore *not* real!

The upper limit to the SSFR at $10^{-8}$ per year is *intrinsic* to our definition of SSFR. The SFR of a model galaxy is defined as the stellar mass it formed in the last $10^8$ years divided by the time elapsed, i.e. $10^8$ years. The maximum value for a SFR according to this definition occurs when the galaxy has formed all of its stars in the last $10^8$ years. If we denote the present stellar mass by $M_{now}$, we therefore obtain for the maximum SSFR:

$$\text{SSFR}_{max} = \frac{M_{now}}{10^8 \text{yr}} / M_{now} = 10^{-8} \text{yr}^{-1}. \quad (3)$$

For comparison purposes we also show literature values derived using primary star formation rate indicators. The most complete comparison is to the data published by Zheng et al. (2007, dotted line), who have redshift bins approximately similar to ours ([0.2,0.4], [0.4,0.6], [0.6,0.8], [0.8,1.0]). They derive the SFR again from observations of the 24 $\mu$m band with Spitzer/MIPS plus an estimate of the rest frame UV flux. Their lowest SFR objects are actually not detected individually in the IR, so they derive their mean IR flux from stacking. There seems to be a small offset in our two lower redshift bins with regard to the Zheng et al. (2007) relation. Whether this indicates that our SFRs are somewhat underestimated, that their sample is somewhat more biased to intrinsically blue galaxies (selected in the R-band with $m_R < 24$) or that the systematic uncertainties in the recovery of the total IR flux from the 24$\mu$m flux alone have been underestimated would warrant a study of its own. Martin et al. (2007b, long-dash-three-dotted line, again in all four redshift bins) have presented the "bivariate M*-SSFR distribution", i.e. what we call the Mass-SFR function. Their SFRs are again based on UV and FIR data. In addition to the SSFR-M* relation, it can also be said that the morphology of the resulting Mass-SFR function agrees well with ours, although we reach two orders of magnitude deeper in stellar mass and we include all galaxies, in particular the quiescent ones. The dot-dashed line in the two upper redshift bins is from Feulner et al. (2005), who derive the SFR from the rest frame UV colour. The two relations we plot here were estimated originally at z=0.6 and z=1.0, respectively, therefore, the lower redshift relation has not been derived from the exact same mean redshift as in our case. Shown in the highest redshift bin only, the dashed line represents the relation between SFR and M* derived by Elbaz et al. (2007) from measurements of the dust emission in the mid-IR (24 $\mu$m observed wavelength) in a redshift interval 0.8-1.2. A figure at least qualitatively similar to Figure 10 has also been shown in Noeske et al. (2007). We conclude that the comparison to the different literature measurements of the SFR is rather good, which we consider additional evidence that SED fitting provides reliable SFRs for large samples of galaxies.

The measurement of the Mass-SFR function bears on many topics that are currently actively studied. Its value as a fundamental measurement to constrain galaxy evolution is underpinned by recent theoretical work, such as e.g. Cattaneo et al. (2007). However, a detailed discussion of the implications of

Figure 10 and a detailed comparison to the literature is beyond the scope of this work. It will be undertaken in forthcoming papers (Lamareille et al., Vergani et al., 2008, in prep.). In the next Section we only concentrate on one possible application.

### 5.3. The evolution of the stellar mass function

We use the Mass-SFR function for a useful test of the merger rate in the universe. To that end we use the Mass-SFR function in an upper redshift bin to predict the mass function in the lower redshift bin. If galaxies grew only through star formation, this predicted mass function should be consistent with the measured one. In Figure 11 we show the observed mass function at each redshift as a solid line. No errorbars are shown, as the formal statistical uncertainties are very small. However, we note explicitly that the mass function we determine is not measured with the same precision as the one published in P07. In particular, the mass function in Figure 11 shows a number of objects with stellar masses higher than $10^{11.8} M_\odot$. Such high stellar masses are not seen in P07 or in the local mass functions of Cole et al. (2001) or Bell et al. (2003). These objects could possibly be Galactic stars or QSOs contaminating our purely photometric sample. Further uncertainties arise in particular at the high-mass end of the mass function, because we are based on photometric redshifts. Objects whose redshift has been strongly overestimated will appear to be intrinsically more luminous than they are in reality. They thus will artificially swell the high-mass end of the mass function. The percentage of catastrophic redshift failures has been well-quantified and is 5.5% for a sample selected at $i'_{AB} \leq 24$ (Ilbert et al. 2006). In our photometric sample ($I_{AB} \leq 25$) this percentage is expected to be somewhat higher. On the other hand, the percentage of objects out of the total sample that have stellar masses above $10^{11.8} M_\odot$ is 4% and 1% for the two bins at $0.37 < z < 0.57$ and $0.57 < z < 0.83$, respectively. We conclude that catastrophic redshift failures make the high-mass end of our stellar mass function uncertain. It is beyond the scope of this paper to remedy this; however, under the assumption that the catastrophic failures redistribute themselves evenly in redshift space, we can draw conclusions about the relative evolution of the mass function at masses below $10^{11.8} M_\odot$, as the fraction of objects from catastrophic failures becomes small below this limit.

Given the stellar mass $M^*_i$ and the SFR $SFR_i$ at time $t$, the stellar mass of galaxy $i$ at time $t + dt$ is simply

$$M^*_i(t + dt) = M^*_i(t) + (\text{SFR}_i(t) \times dt) \times 0.6. \tag{4}$$

The factor 0.6 takes into account the mass loss from stars to the interstellar medium, assuming immediate recycling. To compute the evolution of the mass function self-consistently, one would have to measure the distribution of SFRs in dependence on stellar mass at different redshifts spaced by the typical timescale over which the SFR for a particular galaxy can vary. This timescale, however, is rather short, of the order of $10^8$ years. The present samples are still too small to have significant statistics in each of these small bins. We therefore simplify our approach assuming that:

– The dependence of SFR on stellar mass remains constant over a significant time interval, here 1.5 Gyr. This is not entirely true, as it is known from a number of studies and confirmed here that there is evolution in the star formation activity of galaxies with redshift.
– A galaxy always has a SFR typical of the stellar mass it had at the beginning of the integration. In other words, we assume the typical galaxy will not change its stellar mass significantly over 1.5 Gyr.

**Table 4.** Observed versus predicted evolution in stellar mass density

| Redshift interval (1) | $\Delta\rho^a_{obs}$ (2) | $\Delta\rho^a_{predicted}$ (3) |
|---|---|---|
| from 0.70 to 0.47 | $1.5 \times 10^8$ | $9.7 \times 10^7$ |
| from 1.02 to 0.70 | $9.7 \times 10^7$ | $1.1 \times 10^8$ |

[a] The densities are given in $M_\odot$ Mpc$^{-3}$. The timescale is 1.5 Gyr. No uncertainties are quoted, as the error budget is dominated by unknown systematic uncertainties.

Under these assumptions (see also Drory & Alvarez, 2008, for a similar formalism), we can deduce the stellar mass function $\Phi(M^*, t + dt)$ by replacing $SFR_i$ by the mean SFR for the stellar mass of the galaxy at time $t$, $\langle \text{SFR}(M^*,t) \rangle$. We note that a further caveat concerning this method is our exposure to cosmic variance, i.e. that the galaxy population that we use to predict the mass function comes from an entirely different volume than the galaxy population that we use for the measured mass function.

A useful check on this procedure comes from directly comparing the observed evolution of the total mass density to the predicted one. The evolution in total mass density (summed between the completeness limit in mass given in Section 5.1 and $M^* = 10^{11.8}$ $M_\odot$) $\Delta\rho = \rho(z_0) - \rho(z_1)$ is given in Table 4. The agreement between the predicted and the observed evolution can be deemed satisfactory, given the uncertainties in our mass function. These values can also be compared to the total growth in mass density as measured by Arnouts et al. (2007). From the value given by these authors for the stellar mass growth between z=1.2 and z=0, we infer a mass growth of $7 \times 10^7$ $M_\odot$ Mpc$^{-3}$ for the same total time interval as used in our paper, i.e. 3 Gyr, and the same field, thus minimizing our exposure to cosmic variance. While this is a factor of 3 below our estimate, it must be born in mind that the Arnouts et al. (2007) growth rate value assumes that the growth in the mass density is linear with time. However, it is known and can be seen in their Figure 13 that the mass growth is faster at higher redshift. Thus, for our case, the Arnouts et al. (2007) value is a lower limit, which may alleviate some of the discrepancy. Indeed, the total mass growth between z=1.2 and z=0 as found by Arnouts et al. (2007) is $18 \times 10^7$ $M_\odot$ Mpc$^{-3}$, much closer to our value of $(23 \pm 0.2) \times 10^7$ $M_\odot$ Mpc$^{-3}$.

In Figure 11 we show, besides the observed mass function, the mass function as determined from evolution of the observed Mass-SFR function at higher redshift. We omit the lowest possible redshift bin, as the directly observed stellar mass function suffers from selection effects at the massive end due to limitations in the probed volume. Several effects are noteworthy in Figure 11:

– The completeness limit of the mass function shifts to progressively higher galaxy masses with higher redshift. The effect of this limitation can be seen at the low-mass end of the mass functions.
– Galaxies of stellar masses around $1 - 10 \times 10^{10}$ $M_\odot$ are overproduced.
– Galaxies with masses larger than $1 \times 10^{11}$ $M_\odot$ are underproduced.

Overall, the agreement of the mass function determined here with that from P07, i.e. the reference for the VVDS 02hr field, is satisfactory except at the high-mass end, as discusssed.

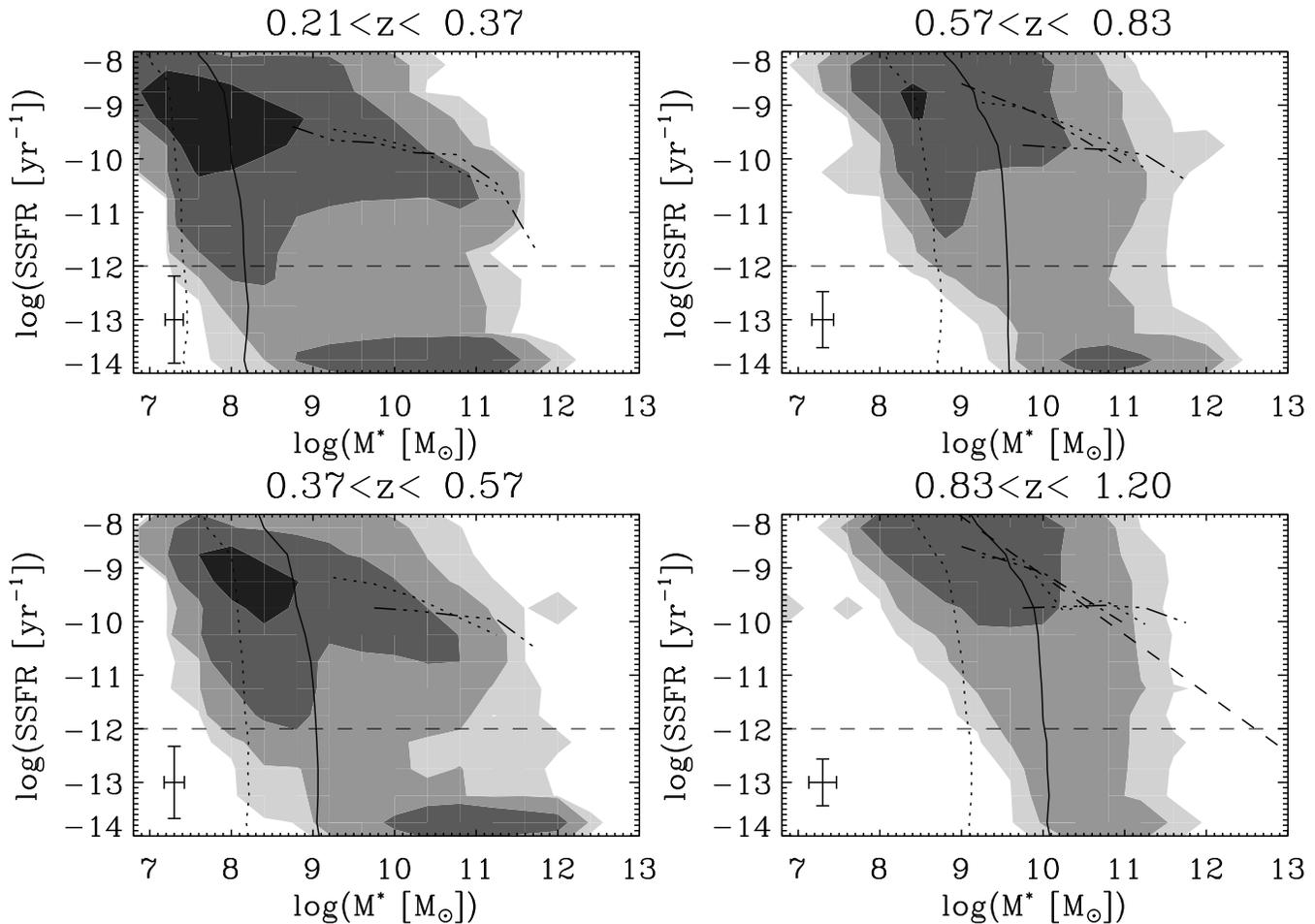

**Fig. 10.** Space density of objects with different M* and SFRs as a function of redshift and as determined from SED fitting. As shown in Section 2.2, the lower limit where we can trust the measurement of the SSFR is at roughly $10^{-12}$ yr$^{-1}$. The horizontal dashed line shows this limit. The objects with measured SSFR below $10^{-14}$ yr$^{-1}$ have been set to $10^{-14}$ yr$^{-1}$. The apparent "sequence" a $10^{-14}$ yr$^{-1}$ is therefore *not* real! The lowest isodensity contour is $10^{-5}$ Mpc$^{-3}$ / log(M*) / log(SSFR) and the contours are spaced by a factor 10. Typical $1\sigma$ errorbars on the measurement for a single object are shown in the lower left corner. The solid and dotted contours show the 10% and 90% completeness limits as derived from our library of models SEDs (see Section 5.1). The solid, dotted and dashed lines show relations between SFR and M* as taken from the literature (see text).

Similar analyses were performed in Bell et al. (2007) and in Martin et al. (2007a,2007b).

Bell et al. (2007) base themselves on SFRs derived from 24 μm data, i.e. a very different SFR tracer. They discuss the evolution of the mass function in a similar way to ours, i.e. predicting the mass function in the lower redshift bin from the mass function and SSFR distribution in the higher redshift bin. They use this approach to ascertain that the growth of the mass in the red sequence of galaxies has to be dominated by the quenching of star formation in blue sequence galaxies. They also state that they find generally good agreement between the predicted and observed total mass functions. Close scrutiny of their Figure 4, however, hints at an opposite trend to the one we find, i.e. the mass function tends to be overpredicted at high masses and underpredicted at intermediate masses. Whether this difference could be attributed to a slight underestimate (overestimate) of stellar masses at lower ( higher) redshifts as mentioned in Bell et al. (2007) remains open.

The comparison to Martin et al. (2007a) will be discussed in the next Section.

### 5.4. Merger rate

We now quantify the evolution in stellar mass. In each bin of the mass function, we compute the following measure of the difference between the predicted mass function $\Phi^{\text{pred}}$ and the originally measured mass functions $\Phi^{\text{orig}}$:

$$\rho_{\text{diff}} = (\Phi^{\text{pred}} - \Phi^{\text{orig}}) * M^*. \quad (5)$$

In those bins where $\rho_{\text{diff}}$ is positive, the growth of the stellar mass as extrapolated from the mass function at higher redshift is larger than the observed value. This means that the simple algorithm we use overpredicts the growth in stellar mass density in this mass bin. We sum this mass into $\rho_{\text{super}}$. In doing this we discard any mass bin below the mass completeness value of the higher redshift bin. In those bins where $\rho_{\text{diff}}$ is negative, the growth of the stellar mass as extrapolated from the mass function at higher redshift is smaller than the observed value, and we sum this mass into $\rho_{\text{under}}$. Due to the relatively high uncertainty on the high-mass end of the mass function, we only consider objects whose stellar mass is below $10^{11.8} M_\odot$. The results are quoted in Table 5.

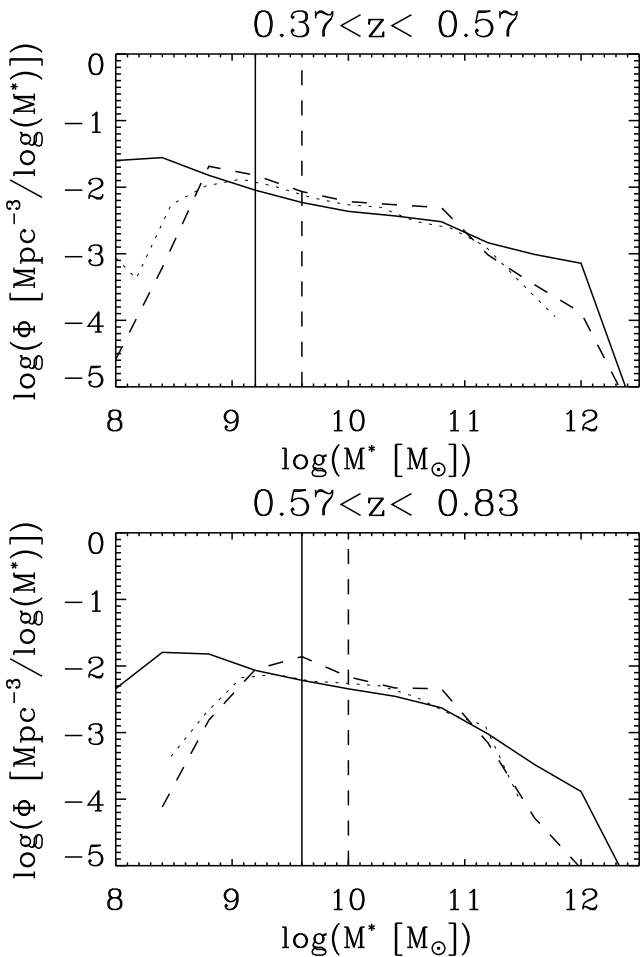

**Fig. 11.** Mass functions as observed and predicted in different redshift bins. The solid line is the observed mass function for the bin shown. The dashed line is the mass function predicted from the higher redshift bin for the current bin. The dotted line is the mass function from Pozzetti et al. (2007). The vertical lines are the corresponding limits of completeness in stellar mass. No errorbars are shown, because the formal statistical uncertainties are very small compared to the systematic uncertainties discussed in the text.

**Table 5.** Stellar mass in mergers as a function of redshift

| Redshift interval | $\Delta\rho^a_{super}$ | $\Delta\rho^a_{under}$ |
| (1) | (2) | (3) |
|---|---|---|
| from 0.70 to 0.47 | $7.7 \times 10^7$ | $1.3 \times 10^8$ |
| from 1.02 to 0.70 | $7.7 \times 10^7$ | $6 \times 10^7$ |

[a] The densities are given in $M_\odot$ Mpc$^{-3}$. The timescale is 1.5 Gyr. No uncertainties are quoted, as the error budget is dominated by unknown systematic uncertainties.

When comparing the predicted and the measured mass functions, it is natural to ask where the mass goes that is overpredicted at intermediate galaxy masses and where the mass comes from that is underpredicted at the highest galaxy masses? A simple physical link between both is provided by merging, i.e. the notion that some or all of the stellar mass that is overpredicted in intermediate mass galaxies must go to the most massive galaxies. A similar argument has been presented e.g. in Brinchman & Ellis (2000). Then, $\rho_{super}$ and $\rho_{under}$ should be equal and provide a direct measure of the stellar mass growth stemming from merger processes. Given the high uncertainties in $\rho_{super}$ and $\rho_{under}$, our results are compatible with this picture. Our best estimate for the stellar mass density involved in mergers between z=1.02 and z=0.47 would then be $(17.2 \pm 3) \times 10^7$ $M_\odot$ Mpc$^{-3}$.

We can directly compare this result with the same quantity measured by another, independent method in de Ravel et al. (2008). These authors have used the spectra of the VVDS to directly measure the fraction of physically associated pairs of galaxies. This can be used to derive a completely independent measure of the stellar mass density that is involved in major mergers. In the redshift bin [1.0,0.4], they derive a value of $11 \times 10^7$ $M_\odot$ Mpc$^{-3}$. The factor 2 difference between both values can be understood when considering that de Ravel et al. (2008) measure the merger rate from galaxies with roughly equal mass ($\Delta M_B \leq 1.5$ mag), i.e. the major merger rate. Taken at face value, this would imply that the contribution of minor and major mergers to the total mass growth of massive galaxies is of similar size.

Martin et al. (2007a) derive quantitative measures for the mass growth in the blue sequence (i.e. through star formation) and for the mass flux to the red sequence (i.e. due to quenching and mergers) of $0.037 \pm 0.006$ $M_\odot$ yr$^{-1}$ Mpc$^{-3}$ and $0.034 \pm 0.031$ $M_\odot$ yr$^{-1}$ Mpc$^{-3}$, respectively. They also note explicitly that these two values are surprisingly close to each other. For a time interval of 3 Gyr, this would be equivalent to a mass flux of $\approx 10^8$ $M_\odot$ Mpc$^{-3}$. However, the identification of massive galaxies with the red sequence and of less massive galaxies with the blue sequence is clearly an oversimplification so we simply conclude with the statement that these two mass fluxes (from intermediate-mass galaxies to high-mass galaxies and from the blue to the red sequence) are comprised within a factor of 2.

That massive galaxies cannot grow by star formation, but must grow through mergers, has also been found by Drory & Alvarez (2008), who follow a very similar line of argument as the one presented here.

## 6. Conclusions

We have used multi-wavelength data from the Vimos VLT Deep Survey and the Galex, CFHTLS and Swire surveys to derive the physical properties of a sample of 84073 galaxies at redshifts between 0 and 1.2. We used a technique in which we use $e^{-\chi^2}$ as a measure of probability and thus build the probability distribution function over parameters such as stellar mass, star formation rate and attenuation. This procedure yields a best estimate of every physical parameter, plus associated errorbars.

We have shown that we can derive accurate measures of $M^*$, SFR and $\langle age_r \rangle$. In particular the measured SFRs are not based on auxiliary data such as the FIR emission or emission lines and are thus per definition internally consistent with the other physical parameters. From independent measurements of the star formation rate as derived from the [OII] line emission we were able to independently test our derived SFR values and we found good agreement.

We have used our catalogue of physical properties to present the number density of galaxies as a function of SSFR and $M^*$. It is particularly convenient that we depend only on one, easily understood selection criterion, i.e. $I_{AB} < 25$. There is no other threshold in a specific emission line or UV / IR photometric band.

We finally used the measured $M^*$ and SFRs in four different redshift bins to predict the stellar mass function that would be observed at a later cosmic time. We find that the predicted evolution in the total mass density agrees with the observed one within the uncertainties. However, star formation alone cannot account for the evolution in the shape of the mass function, as the stellar mass density of intermediate mass galaxies is overpredicted, whereas the density of the most massive galaxies is underpredicted. This indicates that objects with intermediate masses must transform into objects with high masses. The most likely physical cause for this is merging. Indeed, when comparing our results with a direct measure of the major merger rate, we find that approximately half of the mass growth is due to major mergers, while the other half is due to minor mergers.

*Acknowledgements.* CJW is supported by the MAGPOP Marie Curie EU Research and Training Network.

This research was developed within the framework of the VVDS consortium. The VLT-VIMOS observations were carried out on guaranteed time (GTO) allocated by the European Southern Observatory (ESO) to the VIRMOS consortium, under a contractual agreement between the Centre National de la Recherche Scientifique of France, heading a consortium of French and Italian institutes, and ESO, to design, manufacture, and test the VIMOS instrument.

Based on observations obtained with MegaPrime/MegaCam, a joint project of CFHT and CEA/DAPNIA, at the Canada-France-Hawaii Telescope (CFHT), which is operated by the National Research Council (NRC) of Canada, the Institut National des Science de l'Univers of the Centre National de la Recherche Scientifique (CNRS) of France, and the University of Hawaii. This work is based in part on data products produced at TERAPIX and the Canadian Astronomy Data Centre as part of the Canada-France-Hawaii Telescope Legacy Survey, a collaborative project of NRC and CNRS.

[1] Institut d'Astrophysique de Paris, CNRS, Université Pierre & Marie Curie, UMR 7095, 98 bis Boulevard Arago, 75014 Paris, France
[2] Laboratoire d'Astrophysique de Marseille (UMR6110), CNRS-Université de Provence, 38 rue Frederic Joliot-Curie, F-13388 Marseille Cedex 13
[3] Canada France Hawaii Telescope corporation, Mamalahoa Hwy, Kamuela, HI-96743, USA
[4] Centro de Astrofísica da Universidade do Porto, Rua das Estrelas, P-4150-762, Porto, Portugal
[5] INAF-Osservatorio Astronomico di Bologna, Via Ranzani 1, I-40127, Bologna, Italy
[6] IASF-INAF, Via Bassini 15, I-20133, Milano, Italy
[7] IRA-INAF, Via Gobetti 101, I-40129, Bologna, Italy
[8] INAF-Osservatorio Astronomico di Roma, Via di Frascati 33, I-00040, Monte Porzio Catone, Italy
[9] University of California, San Diego 9500 Gilman Dr. La Jolla, CA 92093-0424, USA
[10] California Institute of Technology, MC 405-47, 1200 East California Boulevard, Pasadena, CA 91125
[11] Observatoire de Paris, LERMA, 61 Avenue de l'Observatoire, F-75014, Paris, France
[12] Laboratoire d'Astrophysique de Toulouse-Tarbes, Université de Toulouse,CNRS, 14 avenue Edouard Belin, F-31400 Toulouse, France
[13] School of Physics & Astronomy, University of Nottingham, University Park, Nottingham, NG72RD, UK
[14] Astrophysical Institute Potsdam, An der Sternwarte 16, D-14482, Potsdam, Germany
[15] INAF-Osservatorio Astronomico di Brera, Via Brera 28, I-20021, Milan, Italy
[16] Institute for Astronomy, 2680 Woodlawn Dr., University of Hawaii, Honolulu, Hawaii, 96822, USA
[17] Università di Bologna, Dipartimento di Astronomia, Via Ranzani 1, I-40127, Bologna, Italy
[18] Centre de Physique Théorique, UMR 6207 CNRS-Université de Provence, F-13288, Marseille, France
[19] Integral Science Data Centre, ch. d'Écogia 16, CH-1290, Versoix, Switzerland
[20] Geneva Observatory, ch. des Maillettes 51, CH-1290, Sauverny, Switzerland
[21] The Andrzej Soltan Institute for Nuclear Studies, ul. Hoza 69, 00-681 Warszawa, Poland
[22] INAF-Osservatorio Astronomico di Capodimonte, Via Moiariello 16, I-80131, Napoli, Italy
[23] Leiden Observatory, Leiden University, PO Box 9513, 2300 RA Leiden, the Netherlands
[24] Max-Planck-Institut fur Extraterrestrische Physik, Giessenbachstrasse, D-85748 Garching b. Muenchen, Germany
[25] Universitats-Sternwarte, Scheinerstrasse 1, D-81679 Muenchen, Germany